\documentclass[aps,prd,nofootinbib,twocolumn,superscriptaddress,preprintnumbers,balancelastpage,longbibliography, 10pt]{revtex4-1}

\usepackage{placeins}
\usepackage{amsmath,amssymb,mathtools,bm}
\usepackage{lipsum}
\usepackage{comment}
\usepackage{gensymb}
\usepackage{graphicx, color, hepunits}
\usepackage[dvipsnames]{xcolor}
\usepackage{float}
\usepackage{multirow}
\usepackage{hyperref}
\usepackage{cleveref}
\usepackage{slashed}
\usepackage{hyperref}
\hypersetup{
    colorlinks=true, 
    urlcolor=purple, 
    anchorcolor=blue,
    citecolor=blue, 
    filecolor=magenta,  
    linkcolor=blue, 
    menucolor=blue,
    linktocpage=true,
    pdfproducer=medialab,
}

\usepackage[utf8]{inputenc}
\usepackage[english]{babel}
\usepackage{soul}
\usepackage{hyperref} 
\usepackage{tabularx}
\usepackage{array}
\usepackage{booktabs}
\newcommand{\ra}[1]{\renewcommand{\arraystretch}{#1}}
\makeatletter
\def\hlinewd#1{%
\noalign{\ifnum0=`}\fi\hrule \@height #1 \futurelet
\reserved@a\@xhline}
\makeatother

\newcommand{\es}[2] {\begin{equation} \label{#1} \begin{split} #2 \end{split} \end{equation}}

\makeatletter
\newcounter{savesection}
\newcounter{apdxsection}
\renewcommand\appendix{\par
  \setcounter{savesection}{\value{section}}
  \setcounter{section}{\value{apdxsection}}
  \setcounter{subsection}{0}
  \gdef\thesection{\@Alph\c@section}}
\newcommand\unappendix{\par
  \setcounter{apdxsection}{\value{section}}
  \setcounter{section}{\value{savesection}}
  \setcounter{subsection}{0}
  \gdef\thesection{\@arabic\c@section}}
\makeatother

\begin{document}

\title{
Search for Axions in Magnetic White Dwarf Polarization at  \\ Lick and Keck Observatories
}

\author{Joshua N. Benabou}
\thanks{Corresponding authors: J. N. Benabou, C. Dessert, B. R. Safdi}
\email{joshua\_benabou@berkeley.edu}
\affiliation{Berkeley Center for Theoretical Physics, University of California, Berkeley, CA 94720, U.S.A.}
\affiliation{Theoretical Physics Group, Lawrence Berkeley National Laboratory, Berkeley, CA 94720, U.S.A.}
\author{Christopher Dessert}
\email{cdessert@flatironinstitute.org}
\affiliation{Center for Computational Astrophysics, Flatiron Institute, New York, NY 10010, USA}

\author{Kishore C. Patra}
\affiliation{Department of Astronomy, University of California, Berkeley, CA 94720, U.S.A.}
\affiliation{Department of Astronomy and Astrophysics, UCO/Lick Observatory, University of California, 1156 High Street, Santa Cruz, CA 95064, USA}

\author{Thomas G. Brink}
\affiliation{Department of Astronomy, University of California, Berkeley, CA 94720, U.S.A.}

\author{WeiKang Zheng}
\affiliation{Department of Astronomy, University of California, Berkeley, CA 94720, U.S.A.}

\author{Alexei V. Filippenko}
\affiliation{Department of Astronomy, University of California, Berkeley, CA 94720, U.S.A.}

\author{Benjamin R. Safdi}
\email{brsafdi@berkeley.edu}
\affiliation{Berkeley Center for Theoretical Physics, University of California, Berkeley, CA 94720, U.S.A.}
\affiliation{Theoretical Physics Group, Lawrence Berkeley National Laboratory, Berkeley, CA 94720, U.S.A.}

\date{\today}

\begin{abstract}
We present the most sensitive search to date for light axion-like particles with masses below a micro-eV, using spectropolarimetric data collected from the Lick and Keck Observatories. The conversion of optical photons emitted from the surface of a magnetic white dwarf (MWD) into axions in the strong magnetic field around the star induces a nearly wavelength-independent linear polarization in the observed starlight. We analyze the Stokes parameters $(U, Q, I)$ measured with the Kast spectrograph at the Lick Observatory toward the MWDs SDSS J033320+000720 and ZTF J190132+145807, and with the LRISp-ADC instrument at the Keck Observatory toward ZTF J190132+145807, SDSS J002129+150223, and SDSS J100356+053825 to search for this effect. The data show no evidence of axion-induced linear polarization, and we set world-leading constraints on the axion-photon coupling $|g_{a\gamma\gamma}| \lesssim 1.7 \times 10^{-12} \,\mathrm{GeV}^{-1}$ at the $95\%$ confidence level for masses $m_a \lesssim 2 \times 10^{-7}\,\mathrm{eV}$.
\end{abstract}
\maketitle

\begin{figure}[!htb]
    \centering
    \includegraphics[width=0.48\textwidth]{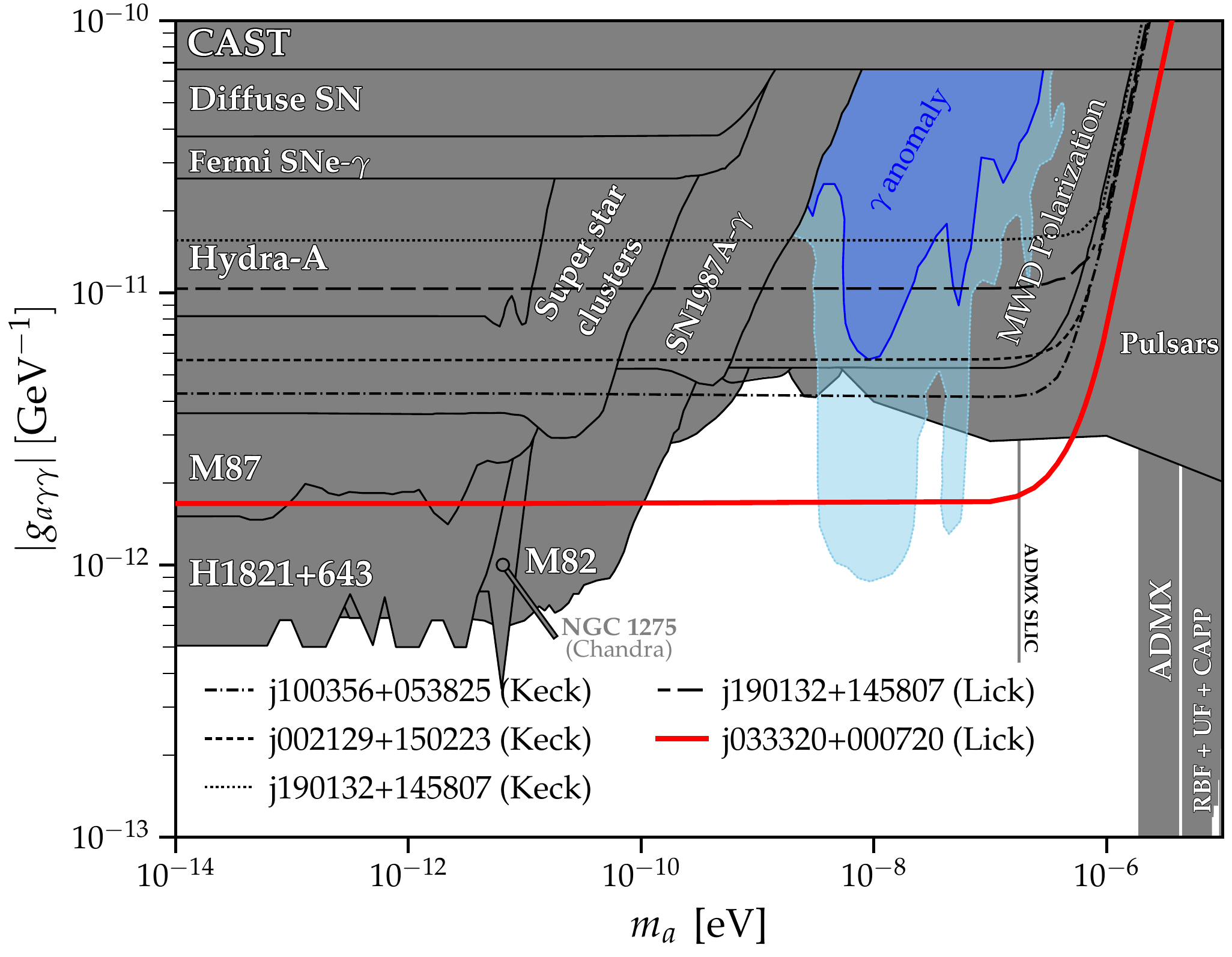}
    \caption{Upper limits on the axion-photon coupling $|g_{a\gamma\gamma}|$ as a function of the axion mass $m_a$ produced in this work from five dedicated spectropolarimetric observations of four MWDs (see Tab.~\ref{tab:results}) for axion-induced linear polarization, with the result from the most constraining MWD in red.
    Existing terrestrial and astrophysical constraints  are shaded in gray (see~\cite{AxionLimits} and text). 
    We also shade the parameter space claimed to be consistent with an axion explanation of the gamma-ray transparency anomaly, for conservative (blue) and optimistic (light blue) assumptions about the intracluster magnetic
field~\cite{Meyer:2013pny}; we disfavor the axion explanation of this anomaly and axion explanations of the GRB 221009A gamma-ray burst.     
    }
\label{fig:limits_summary}
\end{figure}

\textbf{\textit{Introduction}} 
Axion-like particles, which are closely related to the quantum chromodynamics (QCD) axion~\cite{Peccei:1977ur,Peccei:1977hh,Weinberg:1977ma,Wilczek:1977pj}, are strongly motivated hypothetical particles that emerge naturally in  
(for example)
string theory compactifications~\cite{Svrcek:2006yi,Arvanitaki:2009fg}.
Axions couple to electromagnetism through the operator $\mathcal{L} \supset g_{a\gamma\gamma} a {\bf E} \cdot {\bf B}$, with $a$ the axion field, ${\bf E}$ and ${\bf B}$ the electric and magnetic fields (respectively), and $g_{a\gamma\gamma}$ the axion-photon coupling.  The coupling strength $g_{a\gamma\gamma}$ is inversely proportional to the axion decay constant $f_a$, which is the mass scale for the ultraviolet (UV) completion. In string theory UV completions $f_a$ is often related to the string scale, while in field theory UV completions it is tied to the vacuum expectation value of the field that undergoes spontaneous symmetry breaking to give rise to the axion as a pseudo-Goldstone boson (see~\cite{Hook:2018dlk,DiLuzio:2020wdo,Safdi:2022xkm,OHare:2024nmr} for reviews).  QCD axions 
acquire a mass 
$m_a \propto \Lambda_{\rm QCD}^2 / f_a$, with $\Lambda_{\rm QCD}$ the QCD confinement scale; on the other hand, for axion-like particles (referred to simply as axions in this work) there is no direct relation between $g_{a\gamma\gamma}$ and $m_a$, allowing us to consider the full parameter space illustrated in Fig.~\ref{fig:limits_summary}.  Explicit string theory constructions predict axions with $|g_{a\gamma\gamma}|$ only slightly below the current constraints in Fig.~\ref{fig:limits_summary}, strongly motivating more sensitive probes~\cite{Halverson:2019cmy,Gendler:2023kjt} (but see~\cite{Agrawal:2022lsp,Agrawal:2024ejr}).  

In this Letter we search for axions through the axion-photon coupling by studying the conversion of photons emitted from the surface of magnetic white dwarf (MWD) stars into unobserved axions in the strong stellar fields. Initially unpolarized starlight would acquire a linear polarization because the photons polarized perpendicular to the magnetic field are unaffected, while the intensity of light polarized parallel to the magnetic fields is depleted by axion energy loss.  This idea was proposed in~\cite{Lai:2006af,Gill:2011yp,Dessert:2022yqq} but, until this work, no dedicated data have been collected for such a search.  MWDs are not expected to be astrophysical sources of linear polarization, making them ideal environments to search for the axion-induced signal.
In this work we collect optical linear spectropolarimetric data toward some of the most promising MWDs with the Kast double spectrograph mounted on the Shane 3\,m telescope at the Lick Observatory and with the Low Resolution Imaging Spectrometer Polarimeter with Atmospheric Dispersion Compensator (LRISp-ADC) instrument on the Keck-I 10\,m telescope at the W. M. Keck Observatory. We find no evidence for axions and set the strongest constraint to date on $g_{a\gamma\gamma}$ across a broad range of masses. 

In Fig.~\ref{fig:limits_summary} we show existing upper limits on $|g_{a\gamma\gamma}|$ as a function of the axion mass $m_a$~\cite{AxionLimits}, including previous constraints from a MWD polarization search with archival data from the SCORPIO instrument at the Russian Special Astrophysical Observatory 6\,m telescope toward the MWD  SDSS J135141.13+541947.4~\cite{Dessert:2022yqq}.  
All of the leading constraints, except for the haloscope constraints~\cite{Bartram:2024ovw,ADMX:2021nhd,ADMX:2019uok,PhysRevLett.59.839,PhysRevD.42.1297,CAPP:2024dtx,Kim:2023vpo,Yang:2023yry}, are astrophysical and make no dark matter assumption for the axion.
They instead search for relativistic axions produced in stellar cores and converted to photons in laboratory~\cite{CAST:2007jps,CAST:2017uph,CAST:2024eil} or astrophysical~\cite{Dessert:2020lil,Ning:2024eky,Ning:2024ozs,Ning:2025tit,Payez:2014xsa,Manzari:2024jns,Benabou:2024jlj,Noordhuis:2022ljw} magnetic fields, or look for spectral modulations induced by photon disappearance~\cite{Marsh:2017yvc,Conlon:2017qcw,Reynolds:2019uqt,Reynes:2021bpe,Fermi-LAT:2016nkz,Davies:2022wvj,Li:2024zst,MAGIC:2024arq} (though see~\cite{Libanov:2019fzq,Matthews:2022gqi}).

Our new upper limits disfavor axions as an explanation for the observation that the universe appears too transparent to $\sim$TeV gamma-rays~\cite{Dominguez:2011xy,Essey:2011wv,Horns:2012fx,Meyer:2013pny,Rubtsov:2014uga,Kohri:2017ljt}. High-energy gamma-rays from cosmological sources should pair-convert in radiation between us and the source, attenuating the flux. If the gamma-rays convert to axions near the source, however, and then convert back to gamma-rays near the Milky Way, the gamma-ray optical depth is increased.  However, as we show (see, {\it e.g.,}~Fig.~\ref{fig:limits_summary}), the axion parameter space needed to explain these anomalies is strongly constrained by our work.

Axions have similarly been invoked to explain the non-attenuation of high-energy gamma-ray events associated with the gamma-ray burst GRB 221009A~\cite{Galanti:2022chk,Baktash:2022gnf,Lin:2022ocj,Troitsky:2022xso,Zhang:2022zbm,Carenza:2022kjt,Galanti:2022xok,Wang:2023okw,Rojas:2023jdd,Troitsky:2023uwu,Bernal:2023rdz,Gao:2023und}.  The LHAASO air shower array observed an 18 TeV gamma-rays coincident with this event~\cite{LHAASO:2023kyg}, while the Carpet-3 air shower detector may have observed a $\sim$$300$ TeV event~\cite{Dzhappuev:2025ase, Galanti:2025neg, 2022ATel15669....1D}. Ref.~\cite{Baktash:2022gnf} finds $|g_{a\gamma\gamma}| \gtrsim 6 \times 10^{-12}$ GeV$^{-1}$ for $m_a \lesssim 0.1$ $\mu$eV is required if the photons propogated to Earth due to axion conversion, and this parameter space is thoroughly excluded, by a factor of a few in terms of $|g_{a\gamma\gamma}|$, by our search (see Fig.~\ref{fig:limits_summary}).

\begin{table*}[htbp]
\ra{1.3}
\centering
\tabcolsep=0.08cm
\begin{tabularx}{0.86\textwidth}{@{\extracolsep{3pt}}lllcccccc@{}}
\hlinewd{1pt}
& & &  & & & \multicolumn{2}{c}{TS} \vspace{-0.0cm} \\
\cline{7-8}\multicolumn{1}{l}{Magnetic White Dwarf} & \multicolumn{1}{l}{Instrument} & \multicolumn{1}{l}{Date (UTC)} & \multicolumn{1}{c}{$B \, [\mathrm{MG}]$} & 
\multicolumn{1}{c}{$|g_{a\gamma\gamma}^{95}|\ [\mathrm{GeV}^{-1}]$} & \multicolumn{1}{c}{$L_p$ [\%]} & \multicolumn{1}{c}{Astro.} & \multicolumn{1}{c}{Axion} \\ 
\hlinewd{0.5pt}
\href{https://simbad.u-strasbg.fr/simbad/sim-id?Ident=HE\%200330-0002}{SDSS J033320+000720} & Kast & 2023-11-09 & $520^{+319}_{-220}$  & $1.7\times 10^{-12}$ & 0.03  \(\pm\)  0.07 & 0.55 & 0.10 \\ 
\href{https://simbad.u-strasbg.fr/simbad/sim-id?Ident=GALEX\%20J100356.3\%2B053826}{SDSS J100356+053825} & LRISp-ADC & 2023-12-13 & $675_{-248}^{+149}$  & $4.2\times 10^{-12}$ & 0.60  \(\pm\) 0.16 & 5.2 & 7.0 \\
\href{https://simbad.u-strasbg.fr/simbad/sim-id?Ident=SDSS\%20J002128.59\%2B150223.8}{SDSS J002129+150223} & LRISp-ADC & 2023-12-13 & $349_{-93}^{+107}$ & $5.7\times 10^{-12}$ & 0.28  \(\pm\) 0.18 & 6.19 & 1.70 \\
\href{https://simbad.u-strasbg.fr/simbad/sim-id?Ident=IPHAS\%20J190132.77\%2B145807.6}{ZTF J190132+145807} & LRISp-ADC & 2023-06-13 & $565^{+237}_{-198}$ & $1.6\times 10^{-11}$ & 0.33  \(\pm\) 0.17 & 2.45 & 2.47 \\
\href{https://simbad.u-strasbg.fr/simbad/sim-id?Ident=IPHAS\%20J190132.77\%2B145807.6}{ZTF J190132+145807} & Kast & 2023-08-24 & $618_{-246}^{+286}$ & $1.0 \times 10^{-11}$ & 0.30  \(\pm\) 0.06 & 8.6 & 11.9 \\ 
\hlinewd{1pt}
\end{tabularx}
\caption{ 
Analysis results in the $m_a\to 0$ limit for each of the MWD observations performed in this work. For each observation we list the instrument used and the date 
data were obtained.
Additionally, we show results under the harmonic expansion model for the magnetic field geometry with $\ell_\mathrm{max}=2$. From 100 samples of the posterior distribution of magnetic field parameters determined from the flux spectrum fit, we select the sample with the largest likelihood and indicate the corresponding median of the magnetic field strength distribution on the observable hemisphere of the MWD with the $16\%$ ($84\%$) percentiles in sub(super)script. 
Furthermore, from the population of samples we generate an ensemble of upper limits $\{|g^{95}_{a\gamma\gamma}|\}$ at $95\%$ confidence on $|g_{a\gamma\gamma}|$.  
Our fiducial magnetic field configuration is chosen to be that which gives the weakest value of $|g^{95}_{a\gamma\gamma}|$ at $1\sigma$. 
For this configuration we list the 95\% upper limit $|g^{95}_{a\gamma\gamma}|$, the best-fit linear polarization signal amplitude $L_p$ (approximated as a Gaussian-distributed random variable and derived from the wavelength-averaged data),  and the discovery TS for the astrophysical and axion models from the wavelength-dependent and independent analyses, respectively (see text for details). 
\label{tab:results}
}
\end{table*}
\textbf{\textit{Magnetic field model from Zeeman intensity modeling}} 
Our most constraining MWD is J033320+000720 (referred to here as J033320), which we observe with Lick; we describe the analysis toward this MWD in detail below, with details of the other MWDs described afterward and in the Supplementary Materials (SM).  (See also Tab.~\ref{tab:results} for a summary of the results from all of our MWD observations.)  We begin by describing how we determine the magnetic field geometry on the surface of the MWD using spectral modeling. This is important because the axion-induced linear polarization signal scales approximately as the magnetic field strength squared.

\begin{figure}[!t]
    \centering
    \includegraphics[width=0.5\textwidth]{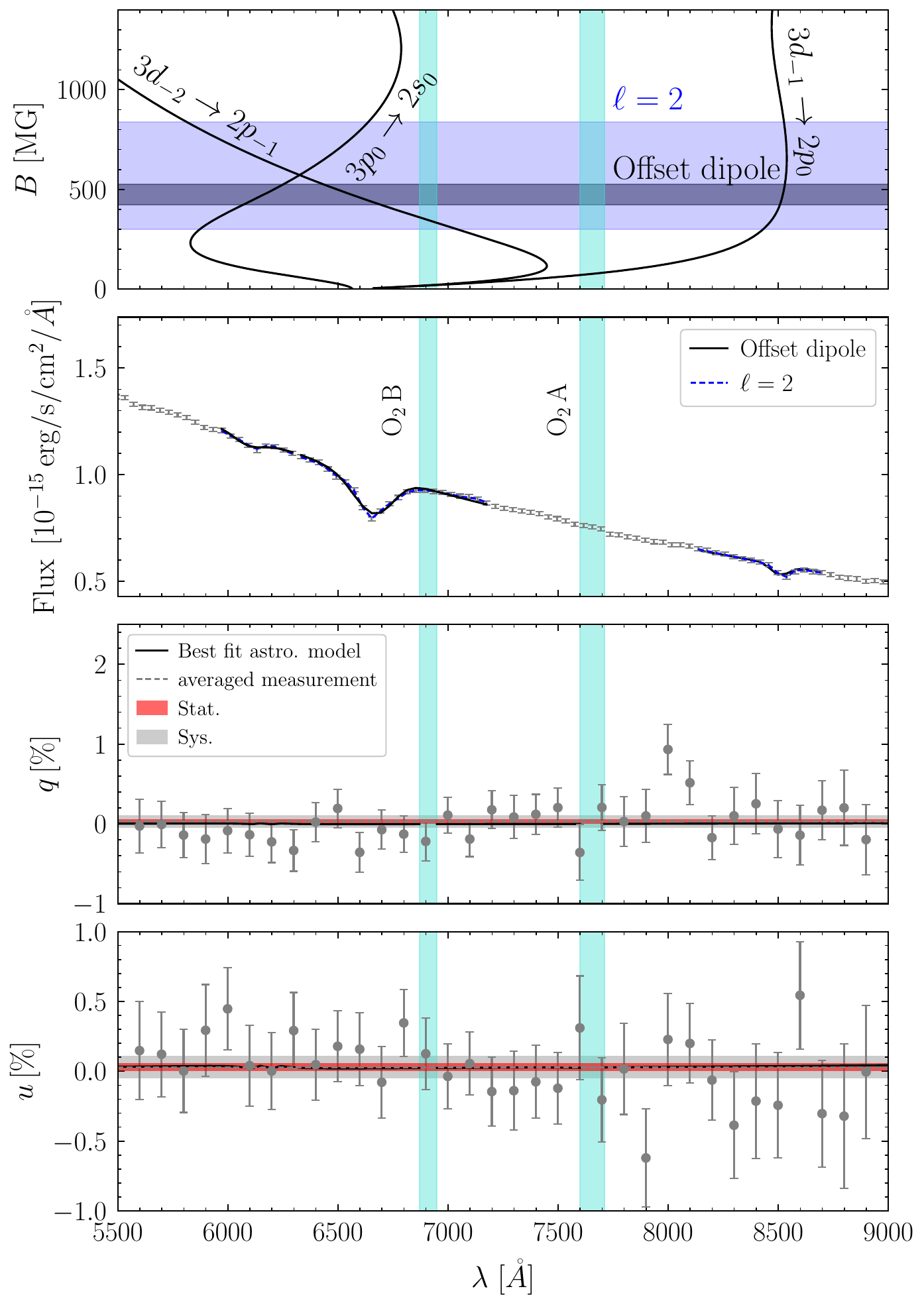}
    \caption{\textit{Top panel}: The wavelengths of the three H$\alpha$ absorption lines considered as a function of magnetic field strength for the J033320 analysis. For the dipole (harmonic) model, we shade in gray (blue) between the $16$th and $84$th percentiles of the magnetic field strength distribution across the observable hemisphere of the star, for a randomly chosen sample in the $68\%$ credible region of the posterior distribution of the magnetic field model parameters. \textit{Middle panel}: The flux measurement from the Kast spectrograph after 20-fold down-binning (gray points), and the fit for a representative offset-dipole ($\ell=2$ harmonic) model chosen from the posterior. 
    \textit{Bottom two panels}: The measured Stokes parameters $u$ and $q$, with $1\sigma$ statistical uncertainties, shown with 50-fold down-binning for clarity. We also indicate the statistical and instrumental systematic uncertainties of the wavelength-independent models for the $q$ and $u$ data (shaded), and the best-fit astrophysical plus systematic model (black), which is consistent with the null hypothesis. In all panels we indicate absorption bands for atmospheric oxygen, which are masked in our work. 
} 
    \label{fig:flux_fit_J033320+00072}
\end{figure}

We self-consistently constrain the magnetic field strength and geometry of each MWD using the spectral intensity data, following the procedure outlined in~\cite{Dessert:2022yqq} based on {\it e.g.}~\cite{Kulebi:2009kp,Euchner:2002qv, Euchner:2006ku}.  
In particular, we fit an astrophysical model  that depends on the magnetic field geometry to the measured intensity spectrum. We take the model to be that of a thermal spectrum with hydrogen absorption lines. For hydrogen-rich stars, only the Balmer-$\alpha$ lines are relevant for the range of magnetic field values and wavelengths considered here.  The key physics point that allows us to determine the magnetic field geometry is that through the quadratic Zeeman effect the line energies are shifted in the presence of magnetic fields by an $\mathcal{O}(1)$ amount (for a review, see~\cite{2000PASP..112..873W}). For a given magnetic field geometry, a range of magnetic field values is visible across the Earth-facing surface of the star, leading to unique spectral features in the absorbed spectra. 

We chose two representative magnetic field models in our work: (i) an inclined and offset magnetic dipole model (3 parameters), and (ii) a harmonic expansion model~\cite{Jordan:2003aa} truncated at $\ell_{\rm max} = 2$ (8 parameters). We use the more general harmonic model as our fiducial model.  Note that in addition to the magnetic field parameters, our models of the spectral intensity data have nuisance parameters for the surface temperature of the star, the overall flux normalization, and for each hydrogen absorption line, the absorption intensity and a Stark broadening parameter. Moreover, the intensity spectral data points are assumed to have uncorrelated and normally distributed uncertainties with common standard deviation $\sigma$, which is treated as a hyperparameter and marginalized over in the analysis.
We apply the limb-darkening law from~\cite{Euchner:2002qv} (see also~\cite{Dessert:2022yqq}) to account for the fact that flux near the star's limb is attenuated relative to flux from the center of the Earth-facing part of the star.  We select wavelengths within the vicinity of relevant transition lines for each of the MWDs and use the spectral intensity data collected by the appropriate instrument ({\it i.e.}, at Keck or Lick). A 20-fold down-binning of the data in wavelength is performed so that each spectral dip is resolved by only a handful of data points to avoid over-resolving the line-like features. We then fit the spectral models to the data using a Bayesian analysis with a Gaussian likelihood. 
Details of these models and the analyses may be found in App.~\ref{app:Bfield}.

In Fig.~\ref{fig:flux_fit_J033320+00072} (second panel from the top) we illustrate the best fits of the magnetospectral models to the spectral data collected toward J033320 by the Kast spectrograph for both the dipole and harmonic models. 
In the top panel we show the field-dependent line wavelengths for the three transition lines in the wavelength range of interest. Note that we select only wavelengths in the illustrated regions to avoid biasing our results from mismodeling thermal emission far away from the absorption lines.  In the top panel we also show the expected range of surface field strengths (at $\pm 1\sigma$) on the observable hemisphere of the MWD in the dipole model and the harmonic model.  Note that we mask wavelengths in the ranges of the shaded molecular oxygen lines, $\sim  [6870,\,6950]$\,\AA\ 
(B band) and $\sim [7600,\,7710]$\,\AA\  
(A band) \cite{1998JGR...10328801N,2022JQSRT.27707949G}, to avoid contamination from Earth's atmosphere. The expected variance (at $\pm 1 \sigma$) on the surface field strengths on the Earth-facing hemisphere of the MWD for the harmonic model are given in Tab.~\ref{tab:results} for each of the observations.  

\textbf{\textit{Search for astrophysical linear polarization}}~In our fiducial analysis we, conservatively, do not account for the possibility of astrophysically induced linear polarization. This is because the axion contribution and the astrophysical contribution to the polarization are expected to add in phase~\cite{Dessert:2022yqq}, so that by neglecting the possibility of astrophysical polarization the upper limits we produce are necessarily weaker than they would be if the astrophysical contribution were accounted for. (On the other hand, this implies we cannot associate a detection of linear polarization to axions without further investigating whether it could be astrophysically produced.)  Still, we begin our analysis by searching for evidence of an astrophysical contribution to the linear polarization. We find no evidence for such a signal from J033320 and at best inconclusive evidence from the other MWDs.  

Our astrophysical spectral model for linear polarization produced in the MWD atmosphere accounts for both bound-free and free-free (cyclotron) emission. 
The free-free transitions introduce polarization at the local surface cyclotron frequency~\cite{lamb_sutherland_1974}, while the bound-free transitions polarize MWD spectra over the continuum~\cite{1992A&A...265..570J}. Additionally, bound-bound transitions polarize the light at the transition wavelength~\cite{1994asmf.book.....R}, though we do not model these transitions in this work. The measured MWD polarization is averaged over the entire observable hemisphere of the star.

The axion-induced linear polarization signal is computed by discretizing the visible hemisphere of the MWD into $N = 64$ pixels, and for each pixel we solve the photon-to-axion mixing equations for light emitted at that point propagating outward toward Earth, for a given magnetic field model. We validate this choice of $N$ by convergence testing.
We model the observed, reduced Stokes parameters $(u,q)=(U/I,Q/I)$ in the wavelength bin $i$, centered at $\lambda_i$, as
\begin{align}
\tilde{s}_i &= s_\mathrm{sys}  + A_\mathrm{astro}s_\mathrm{astro}(\lambda_i,\varphi_m)  \,,
\label{eq:umodel} 
\end{align}
where $s$ denotes $u$ or $q$. 
The model parameter $A_\mathrm{astro}$ is a proxy for the unknown atmospheric depth of the MWD, which rescales the expected strength of the astrophysical linear polarization contribution, and $\varphi_m$ denotes the unknown azimuthal angle of the MWD magnetic field dipole vector with respect to the detector plane. In the absence of circular polarization data, $\varphi_m$ is a priori unknown and is profiled over. 

We also allow for a  systematic offset $u_\mathrm{sys}$ ($q_\mathrm{sys}$) for $u$ ($q$) common to all wavelength bins, with associated systematic uncertainties  \mbox{$\sigma_{\mathrm{sys},u}=\sigma_{\mathrm{sys},q}=\sigma_{\mathrm{sys}}$}.  We take $\sigma_{\mathrm{sys}}=0.08\%$ for the Shane spectrograph and $\sigma_{\mathrm{sys}}=0.21\%$ for LRISp-ADC. (See the SM where we compute these systematic uncertainties from polarization measurements of standard stars that are expected to be unpolarized.)  
Both the Kast and LRISp-ADC data correspond to 1751 wavelength bins between 5500\,\AA\  
and 
 9000\,\AA.   
 The raw Kast data extend to lower wavelengths, but we use a common analysis window to minimize the influence of systematic uncertainties.
 (Note that we again down-bin the data by a factor of 20 for this analysis, as we did for the intensity.)

In the bottom two panels of Fig.~\ref{fig:flux_fit_J033320+00072} we show the $u$ and $q$ data in percent. The best-fit model is indicated, but the astrophysical component is consistent with zero --- the frequentist discovery test statistic (TS) in favor of this model over the null hypothesis of $A_{\rm astro} = 0$ is ${\rm TS} \approx 0.55$ (see Tab.~\ref{tab:results}).  We also show the best-fit $q$ and $u$ values for the constant $u$ and $q$ models over wavelength (labeled ``averaged measurement'') with 1$\sigma$ statistical uncertainties indicated in red. In contrast, the instrumental systematic uncertainties are shown in gray for reference and dominate over the statistical uncertainties.  As shown in Tab.~\ref{tab:results} (with figures in the SM), some of the MWDs show modest evidence in favor of the astrophysical model, though all at less than $\sim$3$\sigma$ significance.  Moreover, it is hard to conclusively say that the linear polarization is arising from the MWD and not another form of terrestrial or astrophysical contamination.  Thus, to prevent setting overly strong constraints, and given our primary MWD shows no signs of polarization whatsoever, we perform our axion search with the wavelength-averaged data, as described below, and neglect the possibility of an astrophysical contribution to the polarization.

\textbf{\textit{ Search for axion-induced linear polarization in wavelength-averaged data}}
The axion-induced polarization signal is proportional to $g_{a\gamma\gamma}^2$ and is approximately independent of $m_a$ for $m_a \ll 10^{-6}$ eV.  At higher axion masses the axion-to-photon dispersion relations are noticeably different over the characteristic distance scale of the MWDs, which suppresses the conversion probabilities. (See App.~\ref{app:axion} for details of this computation.)  The axion-induced linear polarization signal is roughly independent of wavelength over the wavelength range of interest for $m_a \ll 10^{-6}$ eV.  We thus compute the wavelength-averaged Stokes parameters $\bar s = \sum_i (s_i  \sigma_{s_i}^{-2}) / (\sum_i \sigma_{s_i}^{-2})$ with $s$ again denoting $u$ and $q$ and with $\sigma_{s_i}$ the statistical uncertainty in the measurement $s_i$. The uncertainty in $\bar s$ is \mbox{$\sigma_{\bar s} = 1 / \sqrt{ \sum_i \sigma_{s_i}^{-2}}$}.  We then compute the likelihood for the axion model ${\mathcal M}$ with parameter vector ${\bm \theta}$ by
\begin{align}
\mathcal{L}(\mathbf{d} \mid \mathcal{M},\boldsymbol{\theta})&\propto \prod_{s \in \{u,q\}}\left(\frac{1}{\sigma_{\bar s}} e^{\frac{-\left(\bar s-\tilde s(\boldsymbol{\theta})\right)^2}{2 \sigma_{\bar s}^2}} e^{\frac{-s_{\text {sys}}^2}{2 \sigma_{\text {sys},s}^2}} \right) \,.
\label{eq:likelihood_u_q_avg}
\end{align}
Here, the data vector $\mathbf{d}$ consists simply of $\{ \bar u, \bar q\}$, with the associated statistical uncertainties.  The model $\mathcal{M}$ is parameterized by $\boldsymbol{\theta}=\left\{A_{\text {axion }}, \varphi_m, u_{\text {sys }}, q_{\text {sys }}\right\}$ for a fixed MWD magnetic field geometry, and $\tilde s({\bm \theta})$ denotes the axion model plus systematic prediction for the averaged linear polarization for $u$, $q$ (see App.~\ref{app:axion}). 
Note that in~\eqref{eq:likelihood_u_q_avg}, the systematic-uncertainty nuisance parameters, $s_{\rm sys}$, stand for both $u_{\rm sys}$ and $q_{\rm sys}$, with $\sigma_{{\rm sys},s}$ for $s = u,q$ described previously.  The angle $\varphi_m$ describes the orientation angle of the detector plane with respect to the MWD; the $u$ and $q$ values rotate into each other through a unitary transformation under the action generated by $\varphi_m$.  
The signal parameter of interest is $A_{\rm axion}$, which rescales the axion-induced signal and which may be mapped uniquely to $|g_{a\gamma\gamma}|$ for a given MWD magnetic field model and axion mass $m_a$.    We compute the 95\% one-sided upper limit on $A_{\rm axion}$, and thus $|g_{a\gamma\gamma}|$ at fixed $m_a$ and magnetic field model, and search for evidence in favor of the signal model over the null hypothesis ($A_{\rm axion} = 0$) using standard frequentist techniques (see, {\it e.g.},~\cite{Safdi:2022xkm}). 
 As we show through the discovery TSs in Tab.~\ref{tab:results}, none of our analyses find significant evidence for axions, even without accounting for the possibility of astrophysical contributions to the polarization. We thus focus here on the one-sided upper limits on $|g_{a\gamma\gamma}|$ as a function of $m_a$.   

 For a given MWD observation and a given axion mass $m_a$, we compute the upper limit on the axion-photon coupling through the following procedure, accounting for the uncertainties in the MWD magnetic field geometry. First, we draw a large ensemble ($N = 100$) of magnetic field parameter values from the posterior computed in the Bayesian analysis of the intensity data, as described above. Note that we repeat this procedure separately for the dipole and harmonic field models, though our fiducial results use the harmonic model.  For each of the $N$ magnetic field parameters we compute the 95\% one-sided upper limit on $|g_{a\gamma\gamma}|$ at fixed $m_a$ from the frequentist analysis framework of the linear polarization data described above. We then chose our limit to be the weakest limit at 1$\sigma$ ($84^{\rm th}$ upper percentile of the ensemble of upper limits) at a given $m_a$.

\textbf{\textit{Results}} The 95\% one-sided upper limits from our analysis of the J033320 data are illustrated in Fig.~\ref{fig:limits_summary}. As illustrated in Tab.~\ref{tab:results}, the discovery TS in favor of the axion model is insignificant ($\sim$0.7$\sigma$).  At ultra-low axion masses the limit becomes $|g_{a\gamma\gamma}| \lesssim 1.7 \times 10^{-12}$ GeV$^{-1}$.  
In the SM we illustrate the distribution of upper limits that we find varying over the ensemble of $N$ magnetic field model points at low axion mass. The resulting variation in the upper limit is small; for example, the median of the ensemble of upper limits is $|g_{a\gamma\gamma}| \lesssim 1.6 \times 10^{-12}$ GeV$^{-1}$. 
Similarly, using the dipole instead of the harmonic model yields a comparable upper limit of $|g_{a\gamma\gamma}| \lesssim 1.6 \times 10^{-12}$ at low axion masses. (See the SM for further variations.)

We additionally analyze spectropolarimetry data from three other MWDs. This includes ZTF J190132+145807, which is the MWD with the strongest known magnetic field  \cite{Caiazzo:2021xkk} and which we observe with both Kast and LRISp-ADC, as well as SDSS J100356+0538 and SDSS J002129+15022, which we study with LRISp-ADC. The analysis method for these stars is identical to the J033320 analysis.
The Kast data toward ZTF J190132+145807 show marginal evidence for linear polarization, but we note that the Kast observation was taken under cloudy sky conditions.  The data toward  SDSS J100356+05382 also exhibits modest evidence for linear polarization, though at the $\lesssim$3$\sigma$ level (see Tab.~\ref{tab:results}).
(For details of these other MWDs and a discussion of possible polarization contamination from dust, see the SM.)  
We illustrate all of the upper limits as a function of mass in Fig.~\ref{fig:limits_summary}.

\textbf{\textit{Discussion}} 
In this work we perform the most sensitive search to date for low-mass axions but find no evidence for new physics. At the same time, our analyses are limited by instrumental systematics (at least in the case of J033320), which implies that future, improved observations of the targets discussed in this work, and of other promising MWDS, could provide substantially improved sensitivity. 

The FORS2 instrument of the ESO Very Large Telescope could provide such a sensitivity increase, as it can measure linear polarization with a smaller systematic uncertainty than that of the Kast spectrograph --- as low as $0.05\%$ at 4000\,\AA\
and $0.1\%$ at 9000\,\AA~\cite{10.1093/mnras/stw2545}. 
Furthermore, a measurement of the circular polarization of our MWD targets, for example
with Keck-I LRISp-ADC, in conjunction with a modeling of absorption lines in the flux spectra, would permit us to better constrain the magnetic field geometry in the MWD atmosphere~\cite{1992A&A...265..570J,K_lebi_2009}, including determining possible higher-multipole components and the orientation of the detector plane relative to the MWD. 

\textit{ \bf Acknowledgments} --  
We thank Christiane Scherb for collaboration in the early stages of this project, and Joshua Foster and Orion Ning for comments on the manuscript. We are grateful to Sergiy Vasylyev and Yi Yang for support with some of the observations.
J.B. and B.R.S. are supported in part by the DOE award
DESC0025293, and B.R.S. acknowledges support from the
Alfred P. Sloan Foundation.
A major upgrade of the Kast spectrograph on the Shane
3\,m telescope at Lick Observatory, led by Brad Holden, was
made possible through gifts from the Heising-Simons Foundation, William and Marina Kast, and the University of California Observatories. We appreciate the expert assistance of the staff at Lick and Keck Observatories, including Percy Gomez. Research at Lick Observatory is partially supported by a gift from Google. Generous financial support was provided to A.V.F.'s  group by many donors, including the Christopher R. Redlich Fund, Sunil Nagaraj, Landon Noll, Sandy Otellini, Gary and Cynthia Bengier, Clark and Sharon Winslow, Alan Eustace, William Draper, Timothy and Melissa Draper, Briggs and Kathleen Wood, Sanford Robertson 
(K.C.P. was a Nagaraj-Noll-Otellini Graduate Fellow in Astronomy, W.Z. is a Bengier-Winslow-Eustace Specialist in Astronomy, T.G.B. is a Draper-Wood Robertson Specialist in Astronomy).
\\

\bibliography{Bibliography}

\clearpage
\appendix

\section{Data collection}
\label{app:data}

\subsubsection{Lick/Kast data reduction}
\label{sec:Kast_reduction}
In this work, spectropolarimetry of MWDs is conducted using the polarimetry mode of the Kast double spectrograph mounted on the Shane 3\,m telescope at Lick Observatory \cite{miller1994kast}.
In this mode, incoming light passes through a rotatable half-wave plate followed by a Wollaston Prism, which divides the light into two perpendicularly polarized beams, the ordinary and extraordinary beams, displayed as parallel traces on the detector.  
Only the red channel of Kast is used, with a GG455 order-sorting filter that blocks wavelengths below $4550 \, \mathring{A} $ in first order. This also suppresses second-order contamination from light below 4550\,\AA\ 
that would otherwise appear at $< 9100$\,\AA. 
The operational wavelength range spans from 4600 to 9000\,\AA. 
The configuration includes the 300 lines $\mathrm{mm}^{-1}$ grating and a $3''$-wide slit, yielding a spectral resolution of $\sim 18$\,\AA\  
($800 \,\mathrm{km}\, \mathrm{s}^{-1}$)  
at a central wavelength of 6800\,\AA. 
The spectrum is sampled in wavelength bins of size 2\,\AA.

At the beginning of each observation night, flatfield and comparison-lamp exposures are taken. Flatfield spectra are generated using light from an incandescent lamp reflected off the dome's inner surface. Observations of MWDs and calibration standard stars are performed nightly, with four exposures at waveplate angles of $0^{\circ}$, $45^{\circ}$, $22.5^{\circ}$, and $67.5^{\circ}$. The observations are made at low airmass ($\le 1.25$) and the slit is aligned north-south ($180^\circ$).

Instrumental polarization is confirmed to be minimal through nightly observations of the unpolarized standard stars HD~21447 and HD~212311. The instrument's response to $100\%$ polarized light is tested by observing these stars through a polarizing filter. Calibration for accuracy uses nightly spectropolarimetry of high-polarization standard stars chosen among HD~25443, HD~43384, HD~183143, HD~204827, and HD~7927.  The nightly measured Stokes $q$ and $u$ values for the unpolarized standard stars show a consistent low instrumental polarization, which we discuss further in the SM. 
The instrument's polarimetric response, which we measure through polarizance tests, exceeds $99.5\%$ efficiency across the operational wavelength range, precluding the need for additional corrections.

Spectral extraction of the ordinary and extraordinary beams is performed using standard CCD processing and spectral extraction techniques in \texttt{IRAF} \cite{1986SPIE..627..733T}. Images are bias-corrected, and cosmic rays are removed using the \texttt{L.A.Cosmic} algorithm \cite{vanDokkum:2001bp}. Flatfield images are normalized using a low-order spline fit to the continuum, applied to all science and calibration images. Optimal spectrum extraction follows the approach of Ref.~\cite{1986PASP...98..609H}. Wavelength calibration using comparison-lamp exposures is conducted separately for the ordinary and the extraordinary spectra, with fine adjustments based on night-sky emission lines. Flux calibration and telluric absorption correction are performed using a flux standard star observed at similar airmass, fitting splines to the flux-standard spectrum to create a sensitivity function that is subsequently applied to each MWD spectrum. 

Stokes $q$ and $u$ are calculated from two sets of spectra obtained with the waveplate at [$0^{\circ}$, $45^{\circ}$], and [$22^{\circ}$.5, $67^{\circ}$.5], respectively. For each of the ordinary ($o$) and extraordinary ($e$) flux beams, we compute the Stokes parameter
\begin{equation}
q_i= \frac{f_{i,0} - f_{i,45}}{f_{i,0} + f_{i,45}} \,,
\end{equation}
with $i\in \{o,e\}$ and $f_{i,\theta}$ the flux in beam $i$ when the waveplate is at angle $\theta$. Finally, $q$ is obtained as the average of $q_o$ and $q_e$. Similarly, Stokes $u$ is calculated using the exposures at the other set of waveplate positions.

\subsubsection{Keck/LRISp-ADC data reduction}

We use the Keck-I LRISp-ADC instrument \citep{Oke:1995jc,2010SPIE.7735E..0RR} to observe three WD targets at low airmass (in the range $1.03$ to $1.05$).  
Spectropolarimetry is performed using the 400/8500 grating in the red, allowing both moderate velocity resolution and spectral coverage from  
5500\,\AA\ 
to $1\,\mu$m. 
Our default setup employs the $1^{\prime\prime}$ slit. Each linear polarization set consists of 4 exposures with the half-waveplate rotated at angles $0$, $45$, $22.5$, and $67.5$ degrees. The operational wavelength range corresponds to wavelengths between 5500\,\AA\  
and 9000\,\AA,  
where the detector response to $100\%$ polarized light exceeds $99.5\%$.

The data-processing procedures for the Keck data are identical to those used for Lick/Kast. The observed low-polarization standards include BD+284211 and HD~123440, while the high-polarization standards were chosen among BD+64106, HD~236633, HD~155197, and HD~155528. 

Compared with the Kast spectropolarimeter, the LRISp-ADC exhibits a large sensitivity difference between the ordinary (top) and extraordinary (bottom) images captured by  LRISp, which we calculate as  
\begin{equation}
G = \sqrt{\frac{f_{b, 0} ~ f_{b, 45}}{f_{t, 0} ~ f_{t, 45}}} \,,
\end{equation}
where $f_{i,\theta}$ denotes the measured flux from the top ($i=t$) or bottom ($i=b$) of the image, for a waveplate positioned at the angle $\theta$. The Stokes parameter $q_\theta$ when the waveplate is positioned at angle $\theta \in \{0^\circ, 45^\circ\}$ is then 
\begin{equation}
q_\theta = \pm \frac{f_{b, \theta} - G ~ f_{t, \theta}}{f_{b, \theta} + G ~ f_{t, \theta}} \,,
\end{equation}
where ``$+$ is for $\theta = 0^\circ$ and ``$-$'' for $\theta ''= 45^\circ$. Finally, the Stokes parameter $q$ used in the data analysis is the average of $q_0$ and $q_{45}$.
Similarly, the Stokes parameter $u$ is calculated using the exposures at waveplate positions of $22.5^\circ$ and $67.5^\circ$.

\section{Measuring magnetic field strength and geometry}
\label{app:Bfield}

To determine the magnetic field geometry of the observed WD stars, we  fit astrophysical templates to the measured flux spectra.  The astrophysical models consist of a thermal spectrum with hydrogen absorption lines. The absorption lines have field-dependent wavelengths, which allow the magnetic field geometry to be reconstructed. (Note that while the flux spectra are only sensitive to the magnitude of the field, circular polarization spectra --- the subject of future work --- could be used to resolve the orientation of the field relative to the detector plane.) In particular, we perform a fit to the flux spectrum using the transitions $3d_{-2}\to 2p_{-1}$, $3p_0\to 2s_0$, and $3d_{-1}\to 2p_0$. In the case of J100356, the usable wavelength range does not extend past 7800\,\AA. To model the absorption-line profiles, we use two models for the magnetic field geometry: (i) a displaced and misaligned dipole, and (ii) a harmonic expansion model up to and including $\ell = 2$.

Using the dependence of the Balmer-$\alpha$ transition wavelengths and strengths on magnetic field strength, which have been tabulated for $B$ fields up to $1400 \, \mathrm{MG}$~\cite{SCHIMECZEK2014614,DARUS-2118_2021}, we compute the line profiles at a sampling of 12,288 points across the surface of the WD, and then average the profiles over the visible surface of the WD, accounting for limb darkening.  Note that each absorption line is broadened by the Stark effect.  We account for this by smoothing each line profile with a Gaussian kernel with widths independent between lines.

To determine the best-fit model we use a Gaussian likelihood, joint over a selection of wavelengths centered around absorption dips in the data. To mitigate the effect of mismodeling of the absorption dips, the likelihood is applied to the data after down-binning to a bin spacing of 20\,\AA. We omit wavelengths far from absorption dips to reduce systematic errors from mismodeling the background spectrum. The model parameters are
$\boldsymbol{\theta} = \left\{B_0, \iota, r_\mathrm{off},  \sigma, I_i, S_i, N, T_\mathrm{eff} \right\}$. $B_0$ denotes the field strength at the distance $R_\mathrm{WD}$ from the dipole center along the dipole axis, while  $N$ ($T_\mathrm{eff}$) denotes the normalization (effective temperature) of blackbody spectrum,  $dF/d\lambda \propto \lambda^{-5}/\left(e^{hc / (k_B T_\mathrm{eff} \lambda)}-1\right)$. Our derived temperatures may not be consistent with the literature because we fit over a narrow region around the absorption dips and because the blackbody model misses sources of opacity. In this sense $T_{\rm eff}$ functions as a slope parameter rather than the MWD temperature.  $I_i$ is the normalization of the intensity of absorption line $i$, while $S_i$ is the Stark broadening parameter of line $i$.  
We treat the standard deviation $\sigma$ entering in the Gaussian likelihood as a nuisance parameter.  
Note that when exploring the likelihood we do not consider magnetic field geometries for which the WD magnetic field strength exceeds $1400 \, \mathrm{MG}$ at any point on its surface. 
We construct Bayesian confidence intervals by sampling from the posterior distribution via a nested sampling approach. In particular, we use the Python package \texttt{ultranest} \cite{Buchner:2021cql} which implements the \texttt{MLFriends} algorithm with step sampling \cite{Buchner_2019}, parallelized via MPI. For example, in SM Fig.~\ref{fig:posterior_Lick_J033320+00072_ell_2} we show the posterior distribution for our fiducial MWD.

In the offset dipole model we assume flat priors for $B_0$, the inclination angle $\iota$, and the offset of the dipole (in units of the stellar radius) along the inclination direction $r_\mathrm{off}$, and $\sigma$ of $[400,1000]$ MG, $[0^\circ,90^\circ]$, $[-0.5,0.5]$, and $[0,0.2]$ \%, respectively. Log-flat priors are assumed for the parameters characterizing the thermal spectrum, the normalization, and Stark broadening parameters of the absorption features. 
For the harmonic model we expand the magnetic field in the current-free magnetosphere in terms of a magnetic vector potential $\psi$, $\mathbf{B}=-\nabla \psi$, with 
\es{eq:harm}{
\psi=-R_{\mathrm{WD}}\!\sum_{0\le m\le \ell} \! &\left(\!\frac{R_{\mathrm{WD}}}{r}\!\right)^{\ell+1}  \!\!\!P_{\ell}^m \\
&\times \left[g_{\ell}^m \cos m \phi\right.
\left.+h_{\ell}^m \sin m \phi\right] \,,
}
with $P^{m}_\ell(\cos \theta)$ the associated Legendre polynomial and the sum over $\ell \ge 1$. 
Note that formally this parametrization includes the offset-dipole model, though infinitely many harmonics would be required to capture a nonzero offset. Note that here the magnetic field geometry is  parametrized by the $(\ell_\mathrm{max}+1)^2-1$ model parameters $\{g_\ell^m, h_\ell^m\}$. The $h_\ell^0$ are not included to remove a trivial degeneracy. When constructing the posterior distribution for this model (we truncate up to and including $\ell_{\rm max} = 2$), we assume a flat prior for each of the $g_\ell^m, h_\ell^m$ in $[-500,500]$ MG.

\section{Modeling axion-induced linear polarization}
\label{app:axion}

To compute the axion-induced linear polarization, we numerically solve the classical equations of motion of axion electrodynamics for a monochromatic plane wave ansatz as in Ref.~\cite{Dessert:2022yqq} (see \cite{Safdi:2022xkm} for a detailed derivation). The photon wavelengths we probe are much smaller than the extent of the MWD magnetic field, such that a WKB approximation is valid, whereby the second-order system may be reduced to first-order.  We account for the nonlinear Euler-Heisenberg terms, which are relevant for the extreme magnetic fields encountered exterior to the MWDs considered. 

A crucial aspect of the MWD conversion probability calculation is the assumed MWD radius~\cite{Dessert:2022yqq}. In the case of J033320, we infer the stellar radius using the mass measurement from Ref.~\cite{10.1093/mnras/stac3733} and the mass-radius relation from an evolutionary sequence~\cite{2020ApJ...901...93B}, $R_\mathrm{J033320}\approx 1.0 \times 10^{-2}\, R_{\odot}$, while for J190132+145807, we use $R = 0.28 \times 10^{-2}\, R_{\odot}$, as inferred from the mass measurement by Ref.~\cite{Caiazzo:2021xkk}. Note these radii are taken to be those at their lower-$1\sigma$ value. For the MWDs without a measured radius, we assume that they live at the central value of the WD mass distribution, corresponding to $R_\mathrm{WD}=1.12\times10^{-2} R_{\odot}$ \cite{2017ASPC..509..421K}.

\clearpage
\unappendix
\clearpage
\onecolumngrid

\onecolumngrid

\begin{center}
  \textbf{\large Supplementary Material for Search for Axions in Magnetic White Dwarf Polarization at  Lick and Keck Observatories}\\[.2cm]
  \vspace{0.05in}
  {Joshua N. Benabou, Christopher Dessert, Kishore C. Patra, Thomas G. Brink, WeiKang Zheng, Alexei V. Filippenko, and Benjamin R. Safdi}
\end{center}

This Supplementary Material (SM) is organized as follows. In Sec.~\ref{SM:data} we give additional details for the data collection.  In Sec.~\ref{sec:sys_uncertainty} we discuss our inference of the instrumental systematic uncertainties. Section~\ref{SM:ext_fid} gives extended results for our fiducial MWD observation discussed in the main Letter.  In Secs.~\ref{SM:other_1},~\ref{SM:other_2}, and~\ref{SM:other_3} we provide extended details of the other MWD observation analyses.

\section{Extended Details For Spectopolarimetry Data Collection}
\label{SM:data}

In Tab.~\ref{tab:obslog} we provide additional observational details beyond those already in Tab.~\ref{tab:results} in the main Letter.

\begin{table*}[!htb]
\ra{1.3}
  \centering
\tabcolsep=0.08cm
\begin{tabularx}{\textwidth}{@{\extracolsep{1pt}}lllcclclccccc@{}}
\hlinewd{1pt}
  & \multicolumn{2}{c}{Coordinates (J2000)} & & & & $R_{\rm MWD}$ & \multicolumn{2}{c}{Conditions} & \multicolumn{2}{c}{Obs. Info}  \vspace{-0.0cm} \\   
\cline{2-3} \cline{8-9} \cline{10-11} \multicolumn{1}{l}{Magnetic White Dwarf}  & \multicolumn{1}{c}{R.A.} & \multicolumn{1}{c}{Dec.} & $G$ & $E(B-V)$ &  Instrument & $[10^{-2}\, R_\odot]$ & Sky & Airmass &  $N_{\rm loop}$ & $t_{\rm wp}$ \\ 
\hlinewd{0.5pt}
\href{https://simbad.u-strasbg.fr/simbad/sim-id?Ident=HE\%200330-0002}{SDSS J033320+000720} & 03:33:20.37 & +00:07:20.66   & 16.4 & $0.004_{-0.003}^{+0.006}$ & Kast & $1.0$ & Clear  & 1.27  & 3 & 1200 \\
\href{https://simbad.u-strasbg.fr/simbad/sim-id?Ident=GALEX\%20J100356.3\%2B053826}{SDSS J100356+053825}  & 10:03:56.32 & +05:38:25.54  & 18.3 & $0.06_{-0.03}^{+0.02}$ & LRISp-ADC & $1.12$ & Clear & 1.03 & 1 & 1200 \\
\href{https://simbad.u-strasbg.fr/simbad/sim-id?Ident=SDSS\%20J002128.59\%2B150223.8}{SDSS J002129+150223}  & 00:21:28.60 & +15:02:23.91  & 18.0 & $0.03_{-0.02}^{+0.02}$ & LRISp-ADC & $1.12$ & Clear & 1.05 & 1 & 1000 \\
\href{https://simbad.u-strasbg.fr/simbad/sim-id?Ident=IPHAS\%20J190132.77\%2B145807.6}{ZTF J190132+145807}   & 19:01:32.74 & +14:58:07.18 & 15.7 & $0.005_{-0.003}^{+0.010}$ & LRISp-ADC & $0.28$ & Clear & 1.05 & 1 & 400 \\
\href{https://simbad.u-strasbg.fr/simbad/sim-id?Ident=IPHAS\%20J190132.77\%2B145807.6}{ZTF J190132+145807}   & 19:01:32.74 & +14:58:07.18 & 15.7 & $0.005_{-0.003}^{+0.010}$ & Kast & $0.28$ & Cloudy & 1.28  & 2 & 1140 \\
\hlinewd{1pt}
\end{tabularx}
\caption{MWD spectropolarimetry observation log with Kast on the Shane 3\,m telescope at Lick Observatory and LRISp-ADC on the Keck-I 10\,m telescope. For each observation  we indicate the sky coordinates, Gaia $G$ magnitude, color excess $E(B-V)$ in magnitudes of the MWD target, the instrument used, the assumed MWD radii for our analyses (see text), and the conditions when data were obtained. We further provide the number of independent loops, $N_{\rm loop}$, through 4 waveplate position angles as well as the exposure time, $t_{\rm wp}$ in [s], per waveplate position. The total exposure time is $4 N_{\rm loop} t_{\rm wp}$.}
\label{tab:obslog}
\end{table*}

\section{Systematic uncertainty in polarization measurements}
\label{sec:sys_uncertainty}

In this work our sensitivity to axion-induced polarization is limited by systematic uncertainty, with statistical uncertainties being comparatively smaller. We derive our assumed systematic errors in a data-driven way through observations of bright standard stars which are known to be unpolarized. We find consistent results with estimates in the literature. The systematics differ between Kast and LRISp-ADC. 

Systematic errors in $q$ and $u$ from the Kast spectrograph dominantly arise from the fact that to make a single measurement of the Stokes parameters requires four exposures, one for each waveplate angle. For each exposure the extraction and background apertures are set independently, including the ordinary and extraordinary traces. If the airmass or sky conditions change between exposures, the point-spread function varies, which contributes to spurious polarization. This polarization, however, is uncorrelated between $q$ and $u$, and is wavelength-independent. Correlated uncertainties may arise from instrumental miscalibration,  
such as of the zero point of the polarization angle, but which is corrected for.

\begin{figure*}[ht]
    \centering
    \includegraphics[width=0.75\textwidth]
    {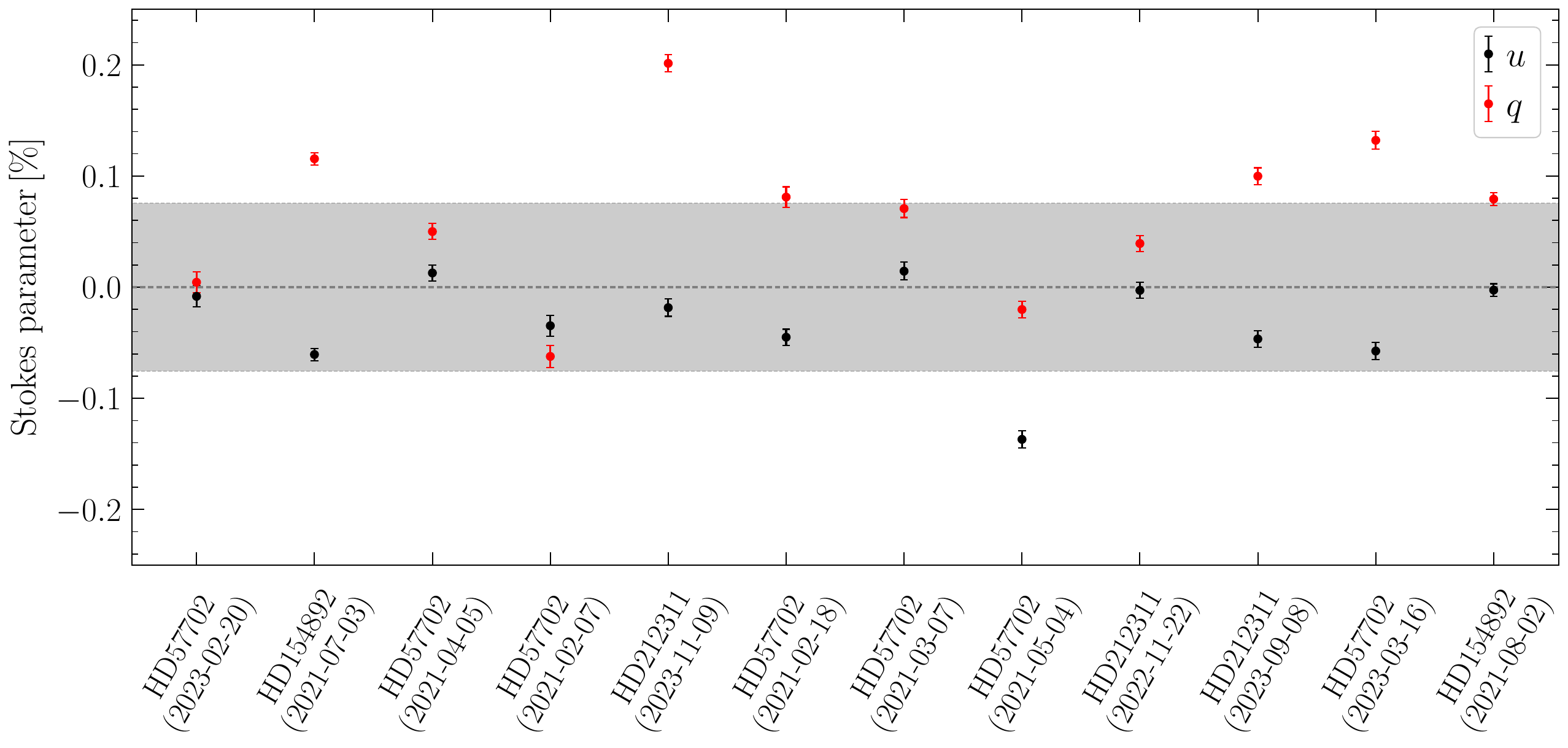}
    \caption{Kast spectrograph polarization measurements from main-sequence unpolarized standard stars (data points, with statistical error bars). The derived systematic uncertainty $\sigma_\mathrm{sys} \approx 0.08$\% is shaded. The observation date is indicated for each measurement.
    }. 
    \label{fig:Lick_systematics_test}
\end{figure*}

To assess the size of the expected spurious polarization in Kast, we analyze polarization data from a set of observations of 12 main-sequence-star observations from the Henry Draper (HD) \href{https://server6.sky-map.org/group?id=23}{catalog}~\cite{vizier:III/135A} with approximately spherical photospheres and which are expected to be unpolarized. In Fig.~\ref{fig:Lick_systematics_test} we show $u$ and $q$ measurements of these 12 observations from the Kast spectrograph, appropriately averaged over our fiducial wavelength range (5500\,\AA\ 
to 9000\,\AA).  
We characterize the data by a joint Gaussian likelihood over $u$ and $q$, and over the 12 observations, with zero mean, and where for $s \in \{q,u\}$ the variance is given by $\sigma_{\mathrm{tot.},q}^2=\sigma_{\mathrm{stat.},q}^2+\sigma_\mathrm{sys}^2$, with $\sigma_{\mathrm{stat.},s}$ the measured statistical uncertainty on $s$ for a given star and $\sigma_\mathrm{sys}$ is the systematic uncertainty, which we assume to be the same for $q$ and $u$. We maximize the likelihood over $\sigma_\mathrm{sys}$ and find the best fit to be $\sigma_\mathrm{sys} \approx 0.08$\%, consistent with the findings of Ref. \cite{Patra:2021ofl}. Note that the statistical uncertainties in the standard-star measurements are generally smaller than in the MWD measurements in this work as the standard stars are significantly brighter, with typical $V \approx 10$--12\,mag compared to the MWDs which are  $V \gtrsim 15$\,mag.

Spurious polarization arises in Keck/LRISp-ADC largely due to stray reflected light within the instrument that happens to fall within the background apertures (see, {\it e.g.},~\cite{2015PASP..127..757H,Goodrich2007}). We estimate its magnitude using observations of four null standard stars:  Gaia EDR3 2504724807045303808 and Gaia EDR3 2504739959689944704, which were observed on 2023-12-13 (UTC) at airmass 1.8 using the LRISp {\it imaging mode} in four broadband filters ($B, V, R, I$) --- which together span the operational wavelength range of the instrument but do not use the ADC component --- and BD+284211 and HD~123440, observed with LRISp-ADC on 2023-12-13 and 2023-06-13 at airmass 1.20 and 1.32 (respectively) using the same set of wavelength bins as in our measurements of the MWDs. 

\begin{figure*}[ht]
    \centering
    \includegraphics[width=0.5\textwidth]
    {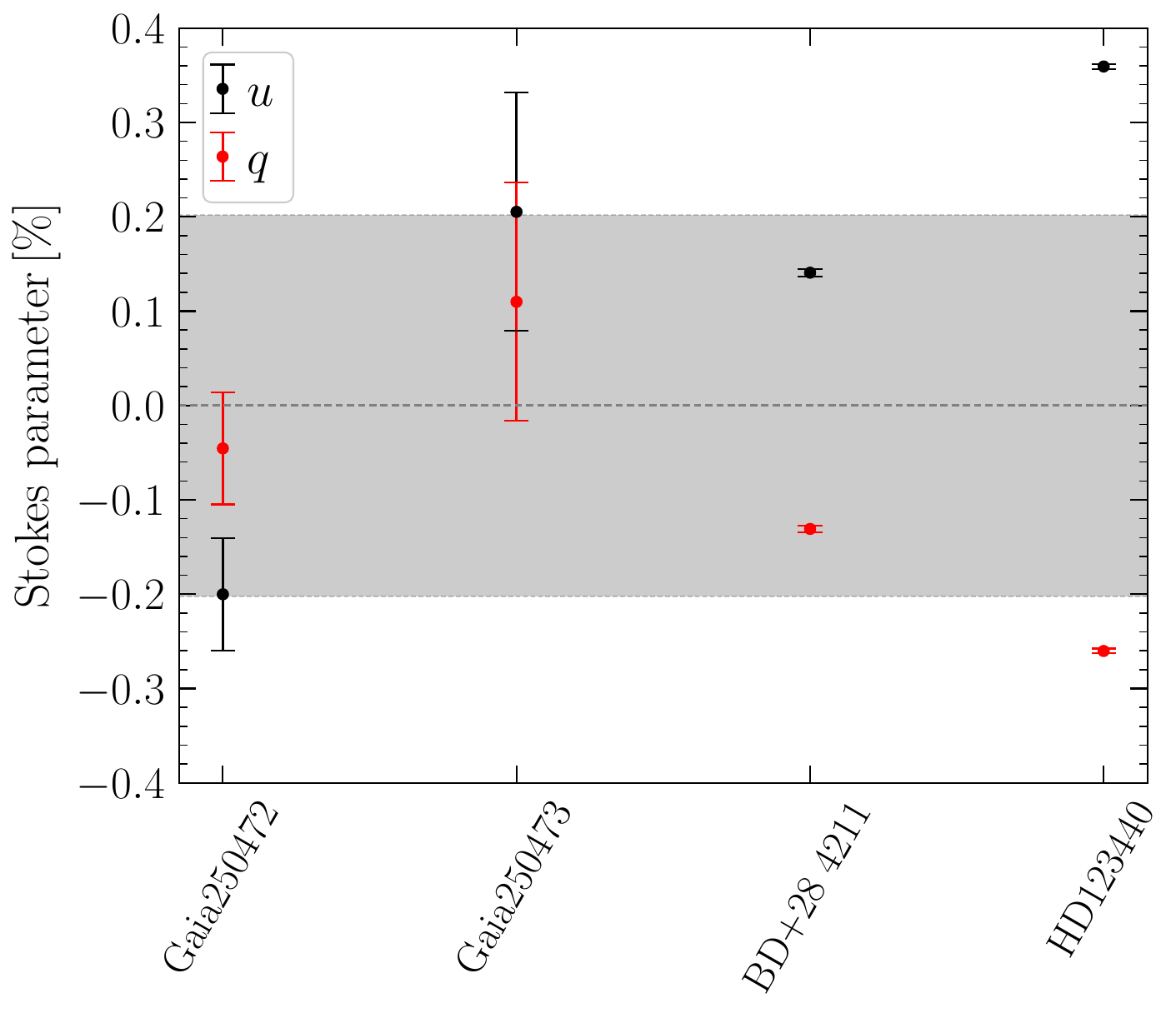}
    \caption{LRISp-ADC polarization measurements from main-sequence unpolarized standard stars (data points, with statistical error bars). The derived systematic uncertainty $\sigma_\mathrm{sys} \approx 0.21$\% is shaded. The Gaia star labels are shortened for clarity.
    }. 
    \label{fig:Keck_systematics_test}
\end{figure*}

These standard stars were observed at relatively high airmass compared to our MWD observations, which are performed close to the zenith. As larger airmasses, the ADC rotates to counter atmospheric dispersion~\cite{1982PASP...94..715F}, which introduces additional systematics that are not expected to be present in the Gaia star measurements (which do not use the ADC), and is unlikely to have an important effect in our MWD observations conducted at low airmass. The measured Stokes parameters for  BD+284211 and HD~123440 show a clear wavelength-dependence, which is likely due to this instrumental artifact. Nonetheless, to be conservative when estimating the systematic uncertainty, we only use the BD+284211 and HD~123440 measurements, from which we derive the best-fit value $\sigma_\mathrm{sys}=0.21\%$. This is very close to the $1\sigma$ upper limit on $\sigma_\mathrm{sys}$ that results from using only the two Gaia star measurements. 
The wavelength-averaged measurements for the four standard stars are shown in Fig.~\ref{fig:Keck_systematics_test}. Note that the statistical errors for BD+284211 and HD~123440 are significantly smaller than for the Gaia star measurements owing to comparatively longer exposures and higher brightnesses.  
A better characterization of the LRISp-ADC systematic uncertainty would be warranted in future work.

Noninstrumental sources of polarization can arise from spheroidal magnetized dust grains present in the interstellar medium along the line of sight. The dust polarization is observed to follow $p_\mathrm{dust} \le 9\%\times E(B-V)$~\cite{Andersson2015}, where $E(B-V)$ is the color excess of the MWD. We note, however, that dust polarization is typically a factor of $\sim2$ lower than the equality. For the nearby J033320, we find $p_\mathrm{dust} \approx 0.02\%$~\cite{2024MNRAS.532.3480D} expected, or $\sim 0.1\%$ at most, with the latter comparable to the systematic uncertainty. Larger dust polarization, up to $\sim 0.5\%$, may contribute in the case of our less-constraining MWDs. 

\section{Extended Details for J033320+00072 Analysis}
\label{SM:ext_fid}

\begin{figure*}[ht]
    \centering
    \includegraphics[width=1.0\textwidth]{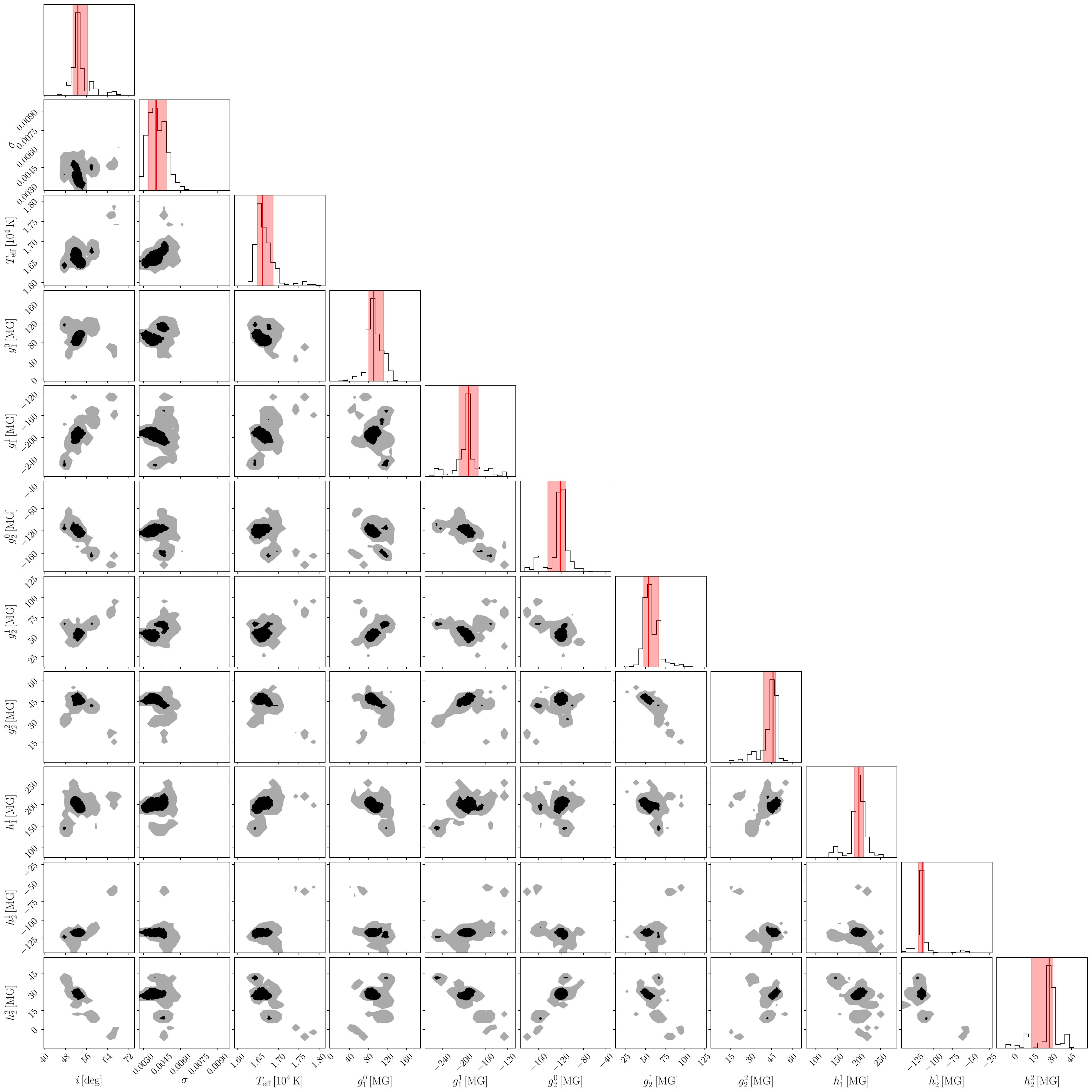} 
    \caption{The posterior distribution of the flux model parameters for the Lick/Kast spectrograph flux measurement of J033320, assuming the    $\ell=2$ harmonic magnetic field model. For brevity, we only show magnetic field parameters and the effective surface temperature of the thermal spectrum (nuisance parameters for the shape of the absorption dips are not shown). The $68.3\%$ ($95.5\%$) credible regions for the joint distributions are shown in black (gray), with the $1\sigma$ confidence interval for each parameter   shaded in red (median indicated by the vertical line).   
    }
    \label{fig:posterior_Lick_J033320+00072_ell_2}
\end{figure*}

\begin{figure*}[ht]
    \centering
    \includegraphics[width=0.8\textwidth]{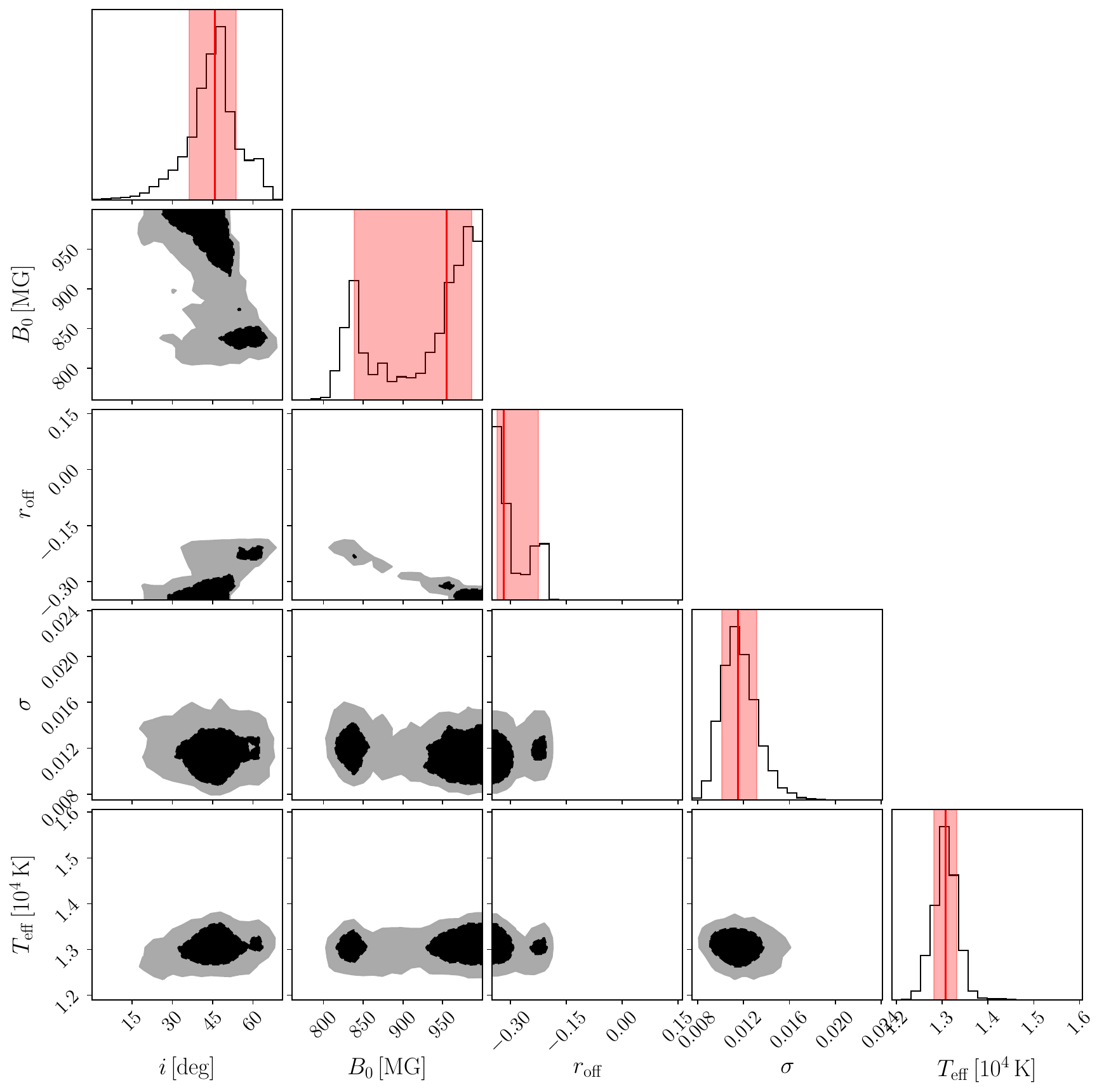} 
    \caption{As in Fig.~\ref{fig:posterior_Lick_J033320+00072_ell_2}, but assuming the dipole magnetic field model.
    }
    \label{fig:posterior_Lick_J033320+00072_offset_dipole}
\end{figure*}

In this section we discuss in more detail our analysis of data collected toward J033320  with the Kast spectrograph. In Fig. \ref{fig:posterior_Lick_J033320+00072_ell_2} (\ref{fig:posterior_Lick_J033320+00072_offset_dipole}) we show the posterior distribution of the magnetic field parameters resulting from the Bayesian analysis of the measured flux spectrum, as described in Appendix \ref{app:Bfield}, assuming the $\ell=2$ harmonic (dipole) model. 

\begin{figure*}[ht]
    \centering
    \includegraphics[width=0.8\textwidth]{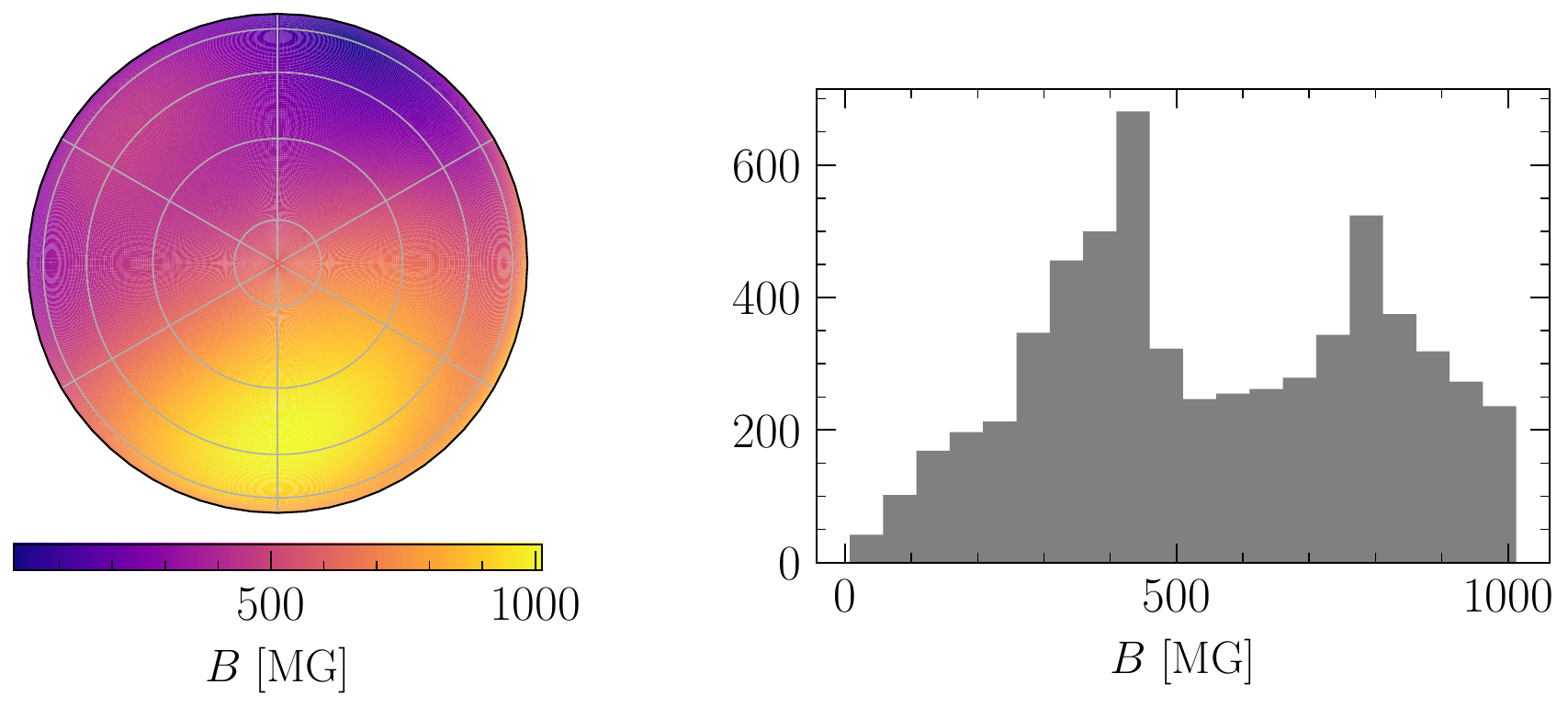} 
    \caption{(\textit{Left}) For the $\ell=2$ harmonic model, the magnetic field strength on the observable hemisphere of J033320 from the sample in the posterior distribution which yields the weakest upper limit on $|g_{a\gamma\gamma}|$ at $1\sigma$.  (\textit{Right}) Histogram of magnetic field strengths across this hemisphere.
    }
    \label{fig:Bfield_dist_ell2_fid}
\end{figure*}

\begin{figure}[h]
    \centering
    \includegraphics[width=1.0\textwidth]{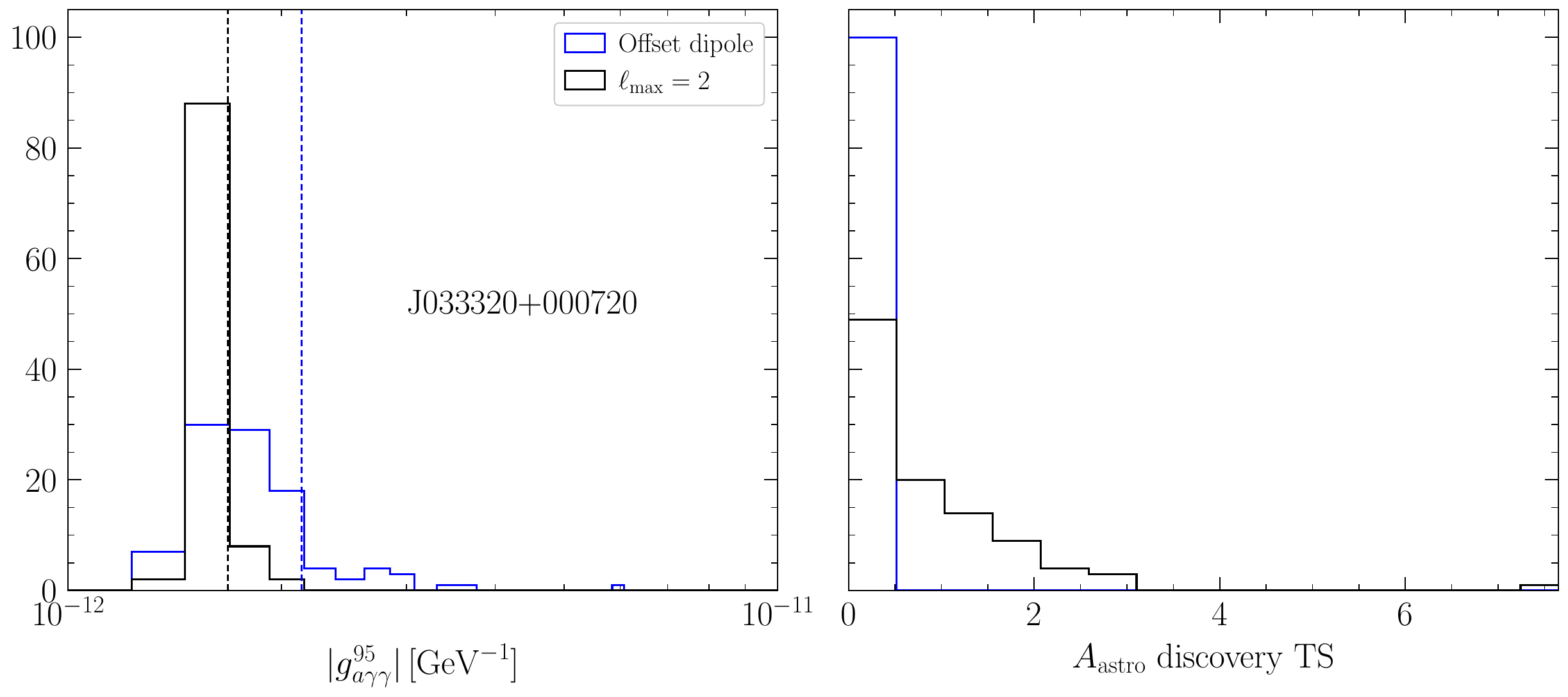} 
    \caption{(\textit{Left}) For the J033320 measurement with the Lick/Kast spectrograph, the distribution of 95\% upper limits on $|g_{a\gamma\gamma}|$ in the $m_a\to 0$ limit for $100$ samples of the posterior distribution of the MWD magnetic field, for the offset-dipole model (blue) and the $\ell=2$ harmonic expansion model (black). For each distribution the 84th percentile is indicated by the dashed line. (\textit{Right}) The distribution of the discovery TS values of the astrophysical component in the wavelength-dependent analysis without axions for each magnetic field model. 
    }
\label{fig:Lick_j033320+000720_3panel_gayy_ul_95_dist_AVG}
\end{figure}

The magnetic field geometry we infer and the effective temperature of the thermal continuum spectrum is broadly consistent with Ref.~\cite{Kulebi:2009kp}. In Fig.~\ref{fig:Bfield_dist_ell2_fid} we show the distribution of the magnetic field strengths over the visible surface of the MWD for the harmonic model, where we take the magnetic field parameter vector to be that which leads to our fiducial limit at asymptotically low axion mass.   
For each magnetic field geometry sampled from the posterior, we perform a profile likelihood on the wavelength-averaged data using a model where the astrophysical component is set to zero, as discussed in the main text. Note that in this analysis, the data are modeled as the sum of the axion signal (computed as described in Appendix \ref{app:axion}) and a systematic uncertainty, which is treated as a nuisance parameter. In the left panel of Fig.~\ref{fig:Lick_j033320+000720_3panel_gayy_ul_95_dist_AVG} we show the resulting distribution of $95\%$ upper limits on $|g_{a\gamma\gamma}|$ (for $m_a=0$) for our fiducial and alternate magnetic field model. In the harmonic magnetic field model, the weakest of these upper limits at $1\sigma$ is $1.68 \times 10^{-12}\, \mathrm{GeV}^{-1}$. To verify the robustness of this result to the modeling of the MWD flux spectrum, we compute the  $95\%$ upper limit on $|g_{a\gamma\gamma}|$ that follows from the findings of Ref.~\cite{Kulebi:2009kp}, which fit a centered-dipole magnetic field model to the SDSS (Sloan Digital Sky Survey) flux spectrum  of J033320. The flux template of Ref.~\cite{Kulebi:2009kp} accounts for radiative transfer effects which we do not include in our modeling. Ref.~\cite{Kulebi:2009kp} found  $B_0=850 \pm 52$\,MG and $\iota= 51^\circ \pm 9^\circ$. We compute a conservative upper limit on $|g_{a\gamma\gamma}|$ by taking the weakest value of $B_0$ and $\iota$ at $1\sigma$, which gives a limit of $1.8 \times 10^{-12}\, \mathrm{GeV}^{-1}$ for $m_a=0$. Note that while this value is weaker than our fiducial limit, the dipole model of Ref.~\cite{Kulebi:2009kp} yields a poor fit to the flux spectrum, and in particular reproduces the position but not the depths of H$\alpha$ transitions, consistent with the findings of this work.

To assess whether the data can be described by our astrophysical model, we compute the distribution of discovery TS values of the astrophysical component, using a joint-likelihood over wavelengths as described in the main Letter, without accounting for a possible axion contribution. This distribution is shown in the right panel of Fig.~\ref{fig:Lick_j033320+000720_3panel_gayy_ul_95_dist_AVG}. Note that in Fig.~\ref{fig:flux_fit_J033320+00072} we show the best-fit astrophysical model in comparison to the measured Stokes parameter for a randomly chosen harmonic magnetic field geometry within the 68\% credible region.

\subsection{Wavelength-dependent axion search for J033320+00072}

As a crosscheck of our fiducial analysis for J033320, we consider a wavelength-dependent search for axions in the Kast polarization data toward this target, profiling both over the systematic nuisance parameters associated with the detector response and, additionally, profiling over the nuisance parameter for the astrophysical contribution to the linear polarization. Recall that this analysis differs from our fiducial analysis method, which only uses the wavelength-averaged data and does not profile over astrophysical contributions to the linear polarization. The analysis method presented here is more aggressive and yields stronger limits, although we feel it is less justified given no convincing evidence that our astrophysical linear polarization model accurately describes the data.

We analyze the data with the likelihood
\es{eq:LL_with_lambda}{
\mathcal{L}(\mathbf{d} \mid \mathcal{M},\boldsymbol{\theta})&\propto \prod_{s \in \{u,q\}} \left(\prod_i \left(\frac{1}{\sigma_{s_i}} e^{\frac{-\left(s_i-\tilde s_i(\boldsymbol{\theta})\right)^2}{2 \sigma_{s_i}^2}} \right) e^{\frac{-s_{\text {sys}}^2}{2 \sigma_{\text {sys},s}^2}} \right) \,,
}
where
\es{}{
\tilde{s}_i &= s_\mathrm{sys}  + A_\mathrm{astro}s_\mathrm{astro}(\lambda_i,\varphi_m) + A_{\rm axion} s_\mathrm{axion}(\lambda_i,\varphi_m) \,.
}
Here, the model ${\mathcal M}$ has two systematic nuisance parameters for $q$ and $u$ data (as before, the astrophysical nuisance parameter $A_{\rm astro}$ and alignment nuisance parameter $\varphi_m$), and then the signal parameter of interest $A_{\rm axion}$, at fixed $m_a$. The function $s_{\rm axion}(\lambda_i,\varphi_m)$ is defined for a reference $|g_{a\gamma\gamma}|$ and a fixed $m_a$. As mentioned previously, and emphasized in~\cite{Dessert:2022yqq}, the functions $s_{\rm astro}$ and $s_{\rm axion}$ are expected to have the same sign, for a given $\lambda_i$ and $\varphi_m$, meaning that they add constructively for positive $A_{\rm astro}$ and $A_{\rm axion}$, which we impose in the optimization. 

To compute the preference for the axion signal model and the 95\% upper limit on $g_{a\gamma\gamma}$ in this analysis, we profile over all nuisance parameters, including $A_{\rm astro}$ while enforcing its positive normalization.   
Applying our standard procedure of choosing the weakest limit from this ensemble at the 84$^{\rm th}$ percentile yields the upper limit $|g_{a\gamma\gamma}| \lesssim 1.66 \times 10^{-12}$ GeV$^{-1}$ at asymptotically low $m_a$, which is slightly stronger than our fiducial upper limit, as expected.

\subsection{Summary of mismodeling systematic uncertainties}

\begin{figure}[!htb]
    \centering
    \includegraphics[width=0.6\textwidth]{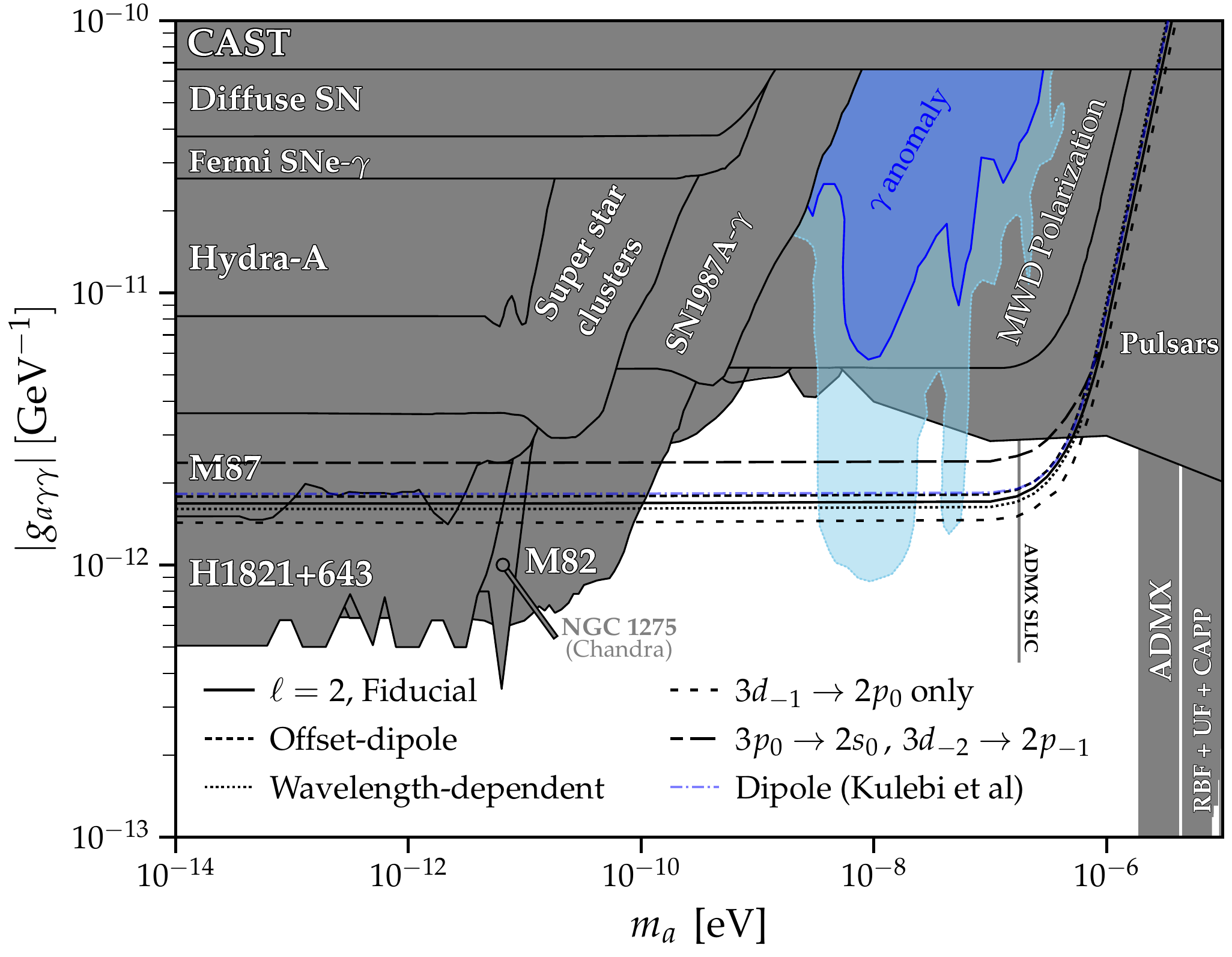}
    \caption{Upper limits on the axion-photon coupling $|g_{a\gamma\gamma}|$ as a function of the axion mass $m_a$ from our Kast spectrograph observation of J033320+00072, for different analysis choices and modeling assumptions. Our fiducial limit, obtained as the weakest limit at $1\sigma$ from the ensemble of magnetic field geometries sampled from the posterior distribution of the $\ell=2$ harmonic model, is shown in solid. 
    Assuming this field geometry, we also show the median of the ensemble of upper limits resulting from our wavelength-dependent analysis (dotted), as well as from two variations of the flux spectrum model, one where only the $3d_{-1} \to 2p_0$ line is included and one where only the $3d_{-2} \to 2p_{-1}$ and $3p_{0} \to 2s_{0}$ lines are included (with line-styles indicated in the legend). Additionally we show the median upper limits obtained from our offset-dipole magnetic field fit (dashed), and a conservative limit using the centered-dipole magnetic field fit of Ref.~\cite{Kulebi:2009kp} (light blue). 
    }
\label{fig:limits_summary_j033_systematics}
\end{figure}

In Fig.~\ref{fig:limits_summary_j033_systematics} we summarize the effects of the systematic uncertainties related to mismodeling the magnetic field and the different analysis choices for our fiducial MWD observation. We shade the band of expected limits at 68\% containment in our fiducial analysis over the ensemble of magnetic field parameter vectors sampled from the Bayesian posterior for the harmonic fit, with the median limit also indicated. These limits are compared to the median limit from the dipole fit and also the limit obtained using the magnetic field model from~\cite{Kulebi:2009kp}. We also show the upper limit obtained from the wavelength-independent analysis described in the previous subsection.

To assess mismodeling arising from our magnetic field measurements, we (in the context of the harmonic model) repeat our analysis procedure using only the $3d_{-1} \to 3 p_0$ transition and then independently using only the two lower-wavelength transitions, instead of the joint analysis over all three transitions as done in our fiducial analysis. We also show the mean limits from these variants in Fig.~\ref{fig:limits_summary_j033_systematics}.

\section{Extended Details for the ZTF J190132+145807 Analysis}
\label{SM:other_1}

We now discuss our analysis results for ZTF J190132+145807, which is the most massive WD known and is rapidly rotating with a period of $6.94$ min \cite{Caiazzo:2021xkk}. This MWD has also been used to search for axions in $X$-rays~\cite{Ning:2024ozs}.
ZTF J190132 is unique among the targets studied in this work because we measure polarimetry of this star using both the Lick/Kast spectrograph and the Keck/LRISp-ADC.
The analysis procedure for ZTF J190132 is the same as for the other MWDs studied in this work, with one key difference. As the exposure time for the polarimetry of this star (on the order of hours) exceeds the rotation period of the MWD, when modeling the polarization and the flux spectrum, we should in principle average the predicted Stokes parameters over the rotational phase of the MWD. (In this work, we do not account for this more sophisticated modeling, though this should be done in future studies.)

\begin{figure}[ht]
    \centering
    \includegraphics[width=0.495\textwidth]{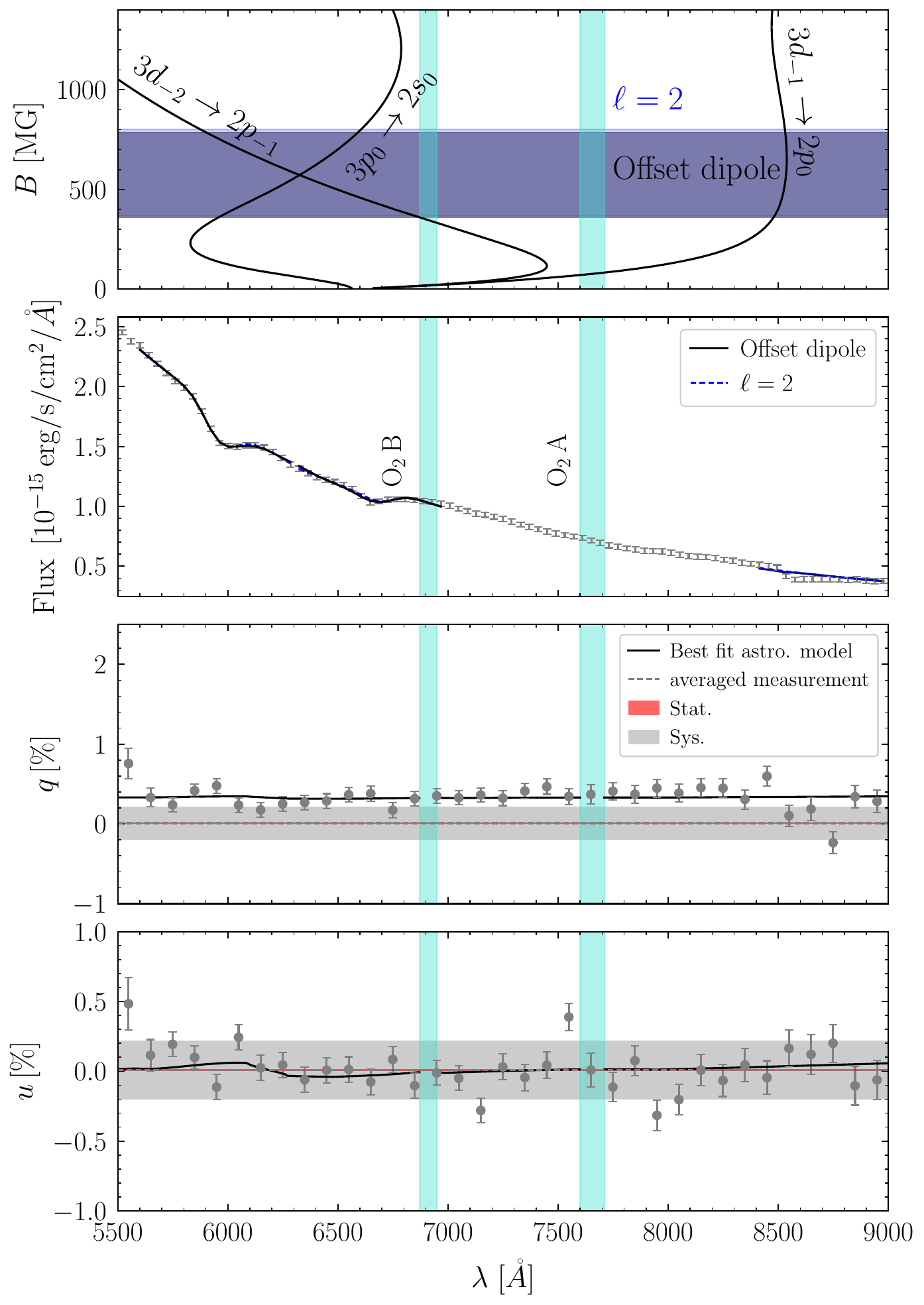}
    \hspace{-2.0em}
    \includegraphics[width=0.495\textwidth]{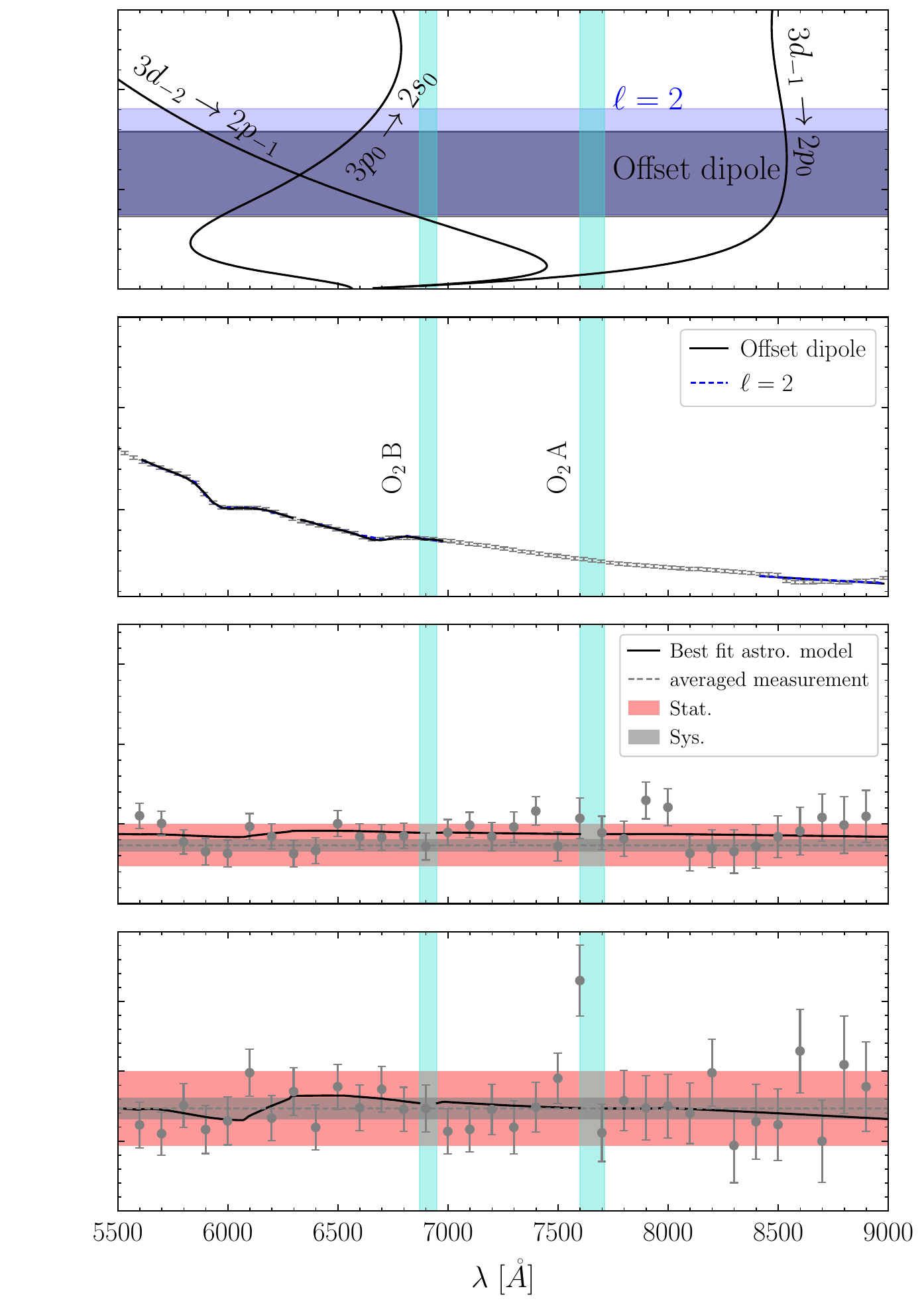}
    \caption{(\textit{Left}) As in Fig.~\ref{fig:flux_fit_J033320+00072} but for the LRISp-ADC observation of J190132+145807. (\textit{Right}) The same, but for the Kast observation of J190132+145807.
    }. 
    \label{fig:flux_fit_j190132+145807_v2}
\end{figure}

\begin{figure*}[ht]
    \centering
    \includegraphics[width=1.0\textwidth]{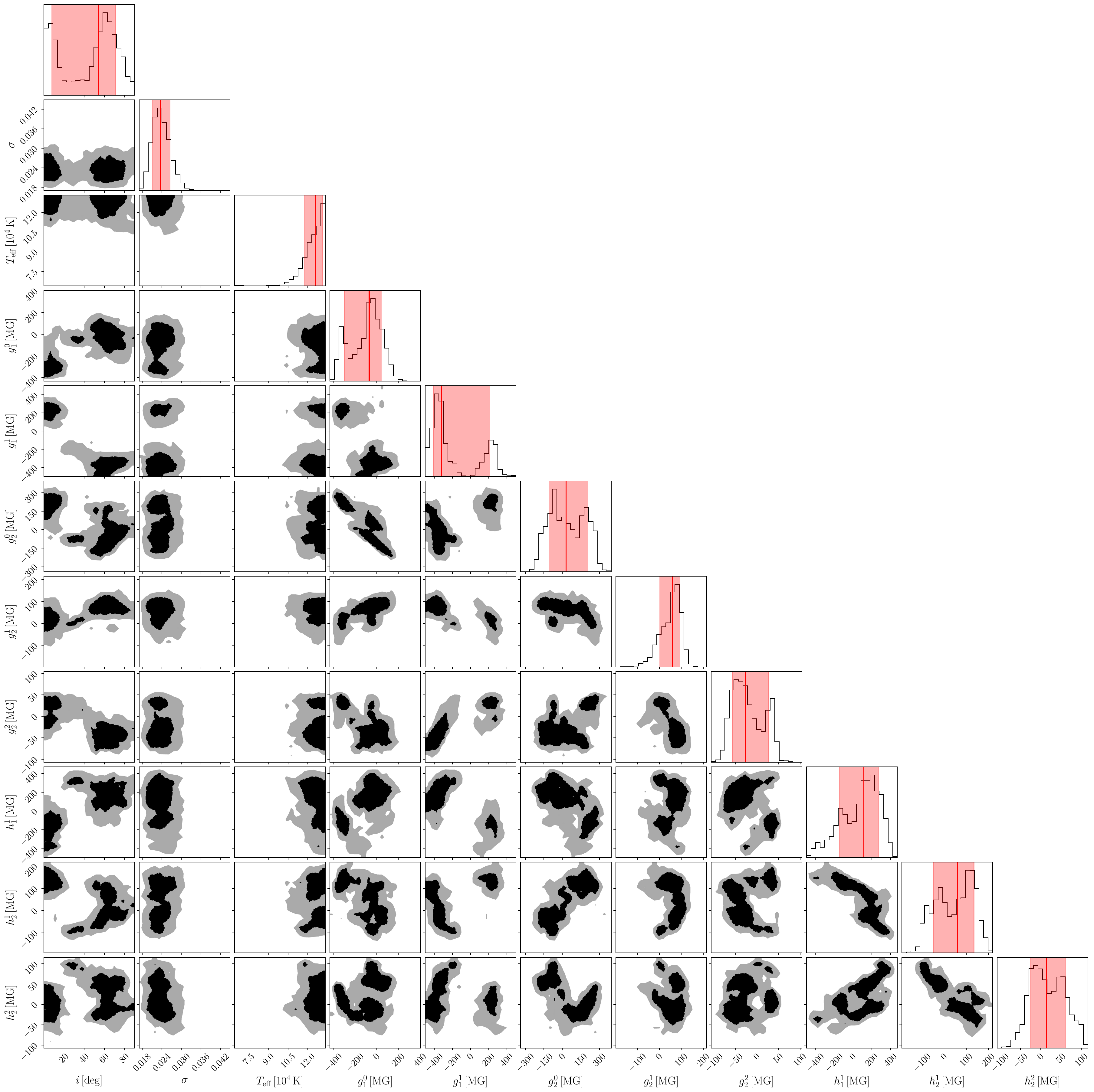} 
    \caption{As in Fig.~\ref{fig:posterior_Lick_J033320+00072_ell_2} but for the Keck/LRISp-ADC measurement of J190132+145807.  
    }
    \label{fig:posterior_Keck_j190132+145807_ell_2}
\end{figure*}

\begin{figure*}[htb!]
    \centering
    \includegraphics[width=1.0\textwidth]{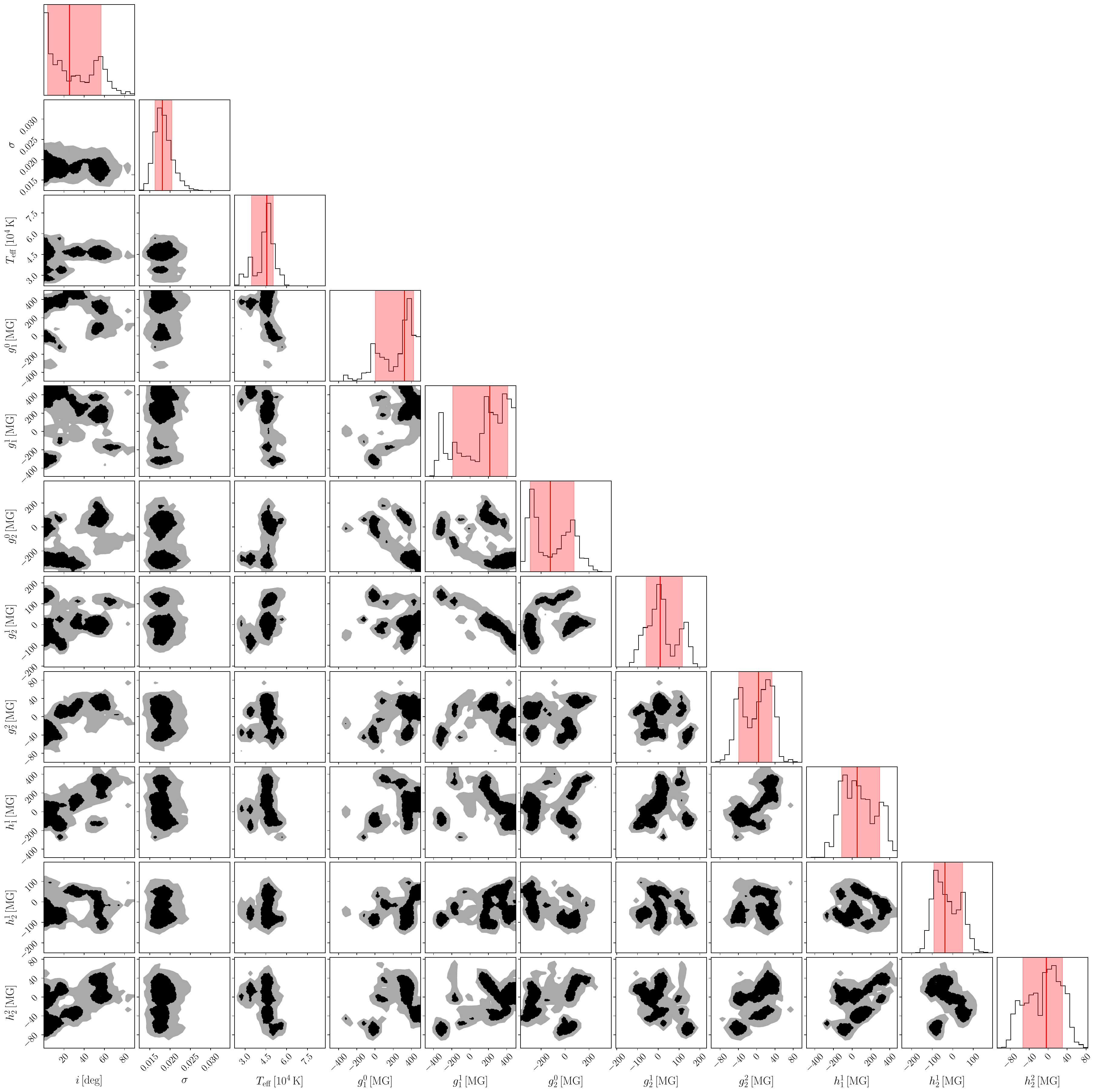} 
    \caption{As in Fig.~\ref{fig:posterior_Lick_J033320+00072_ell_2} but for the Lick/Kast spectrograph measurement of J190132+145807.  
    }
    \label{fig:posterior_Lick_j190132+145807_ell_2}
\end{figure*}

\begin{figure*}[ht]
    \centering
    \includegraphics[width=\textwidth]{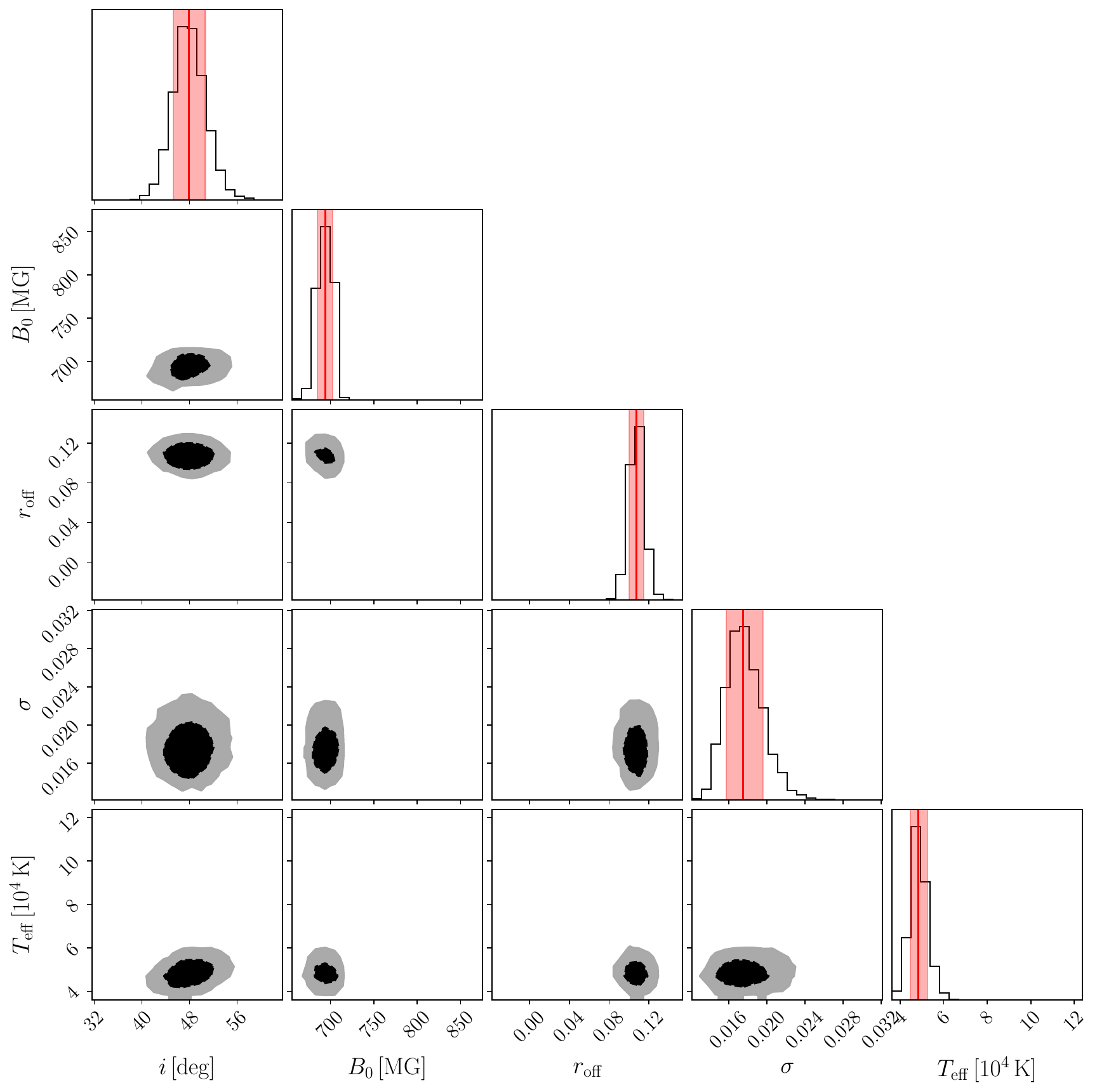} 
    \caption{The posterior distribution in the offset-dipole model for the Lick/Kast spectrograph observation of J190132.
    }
\label{fig:posterior_Lick_J190132_offset_dipole}
\end{figure*}

\begin{figure*}[ht]
    \centering
    \includegraphics[width=\textwidth]{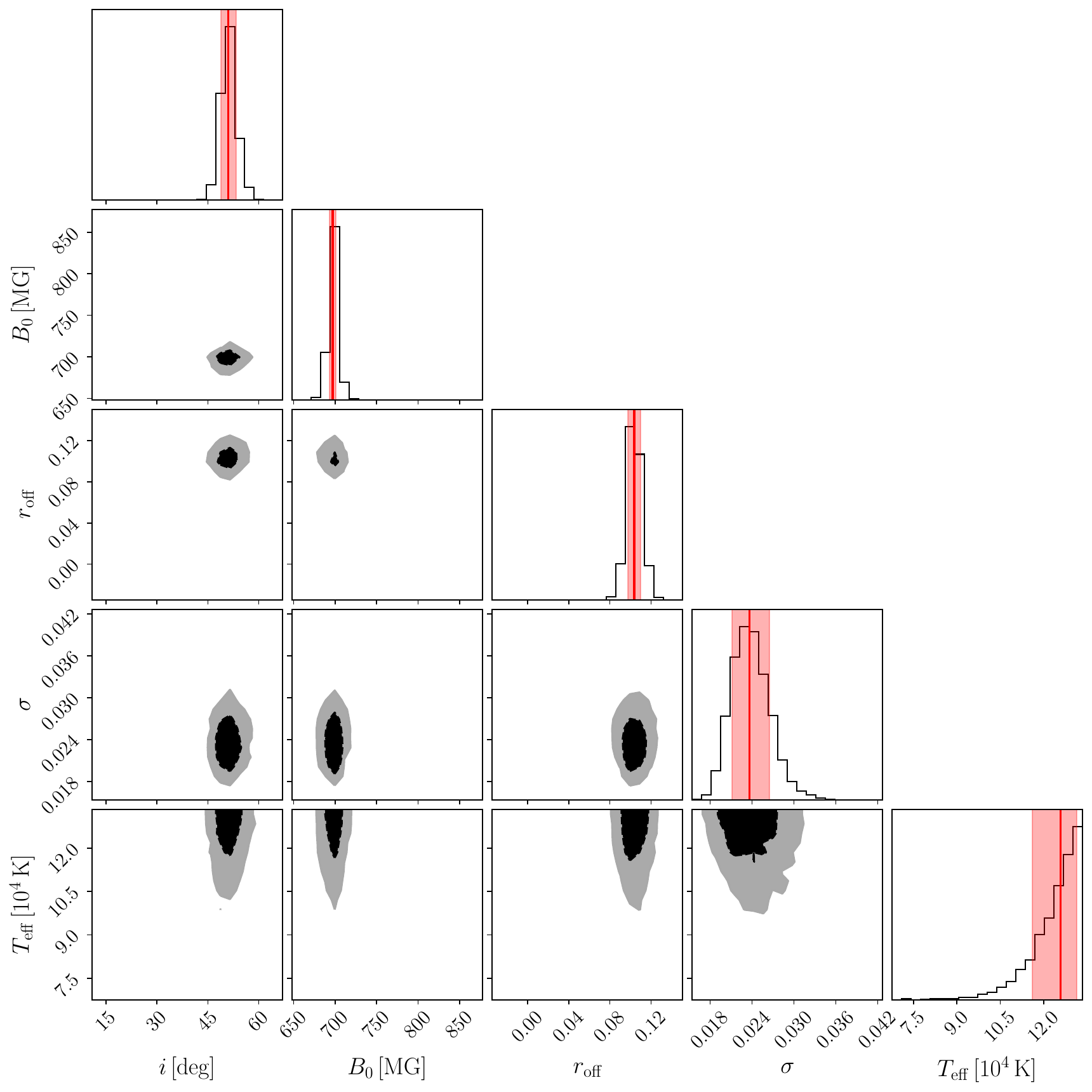} 
    \caption{The posterior distribution in the offset-dipole model for the Keck/LRISp-ADC observation of J190132. 
    }
\label{fig:posterior_Keck_J190132_offset_dipole}
\end{figure*}

\begin{figure}[h]
    \centering
    \includegraphics[width=1.0\textwidth]{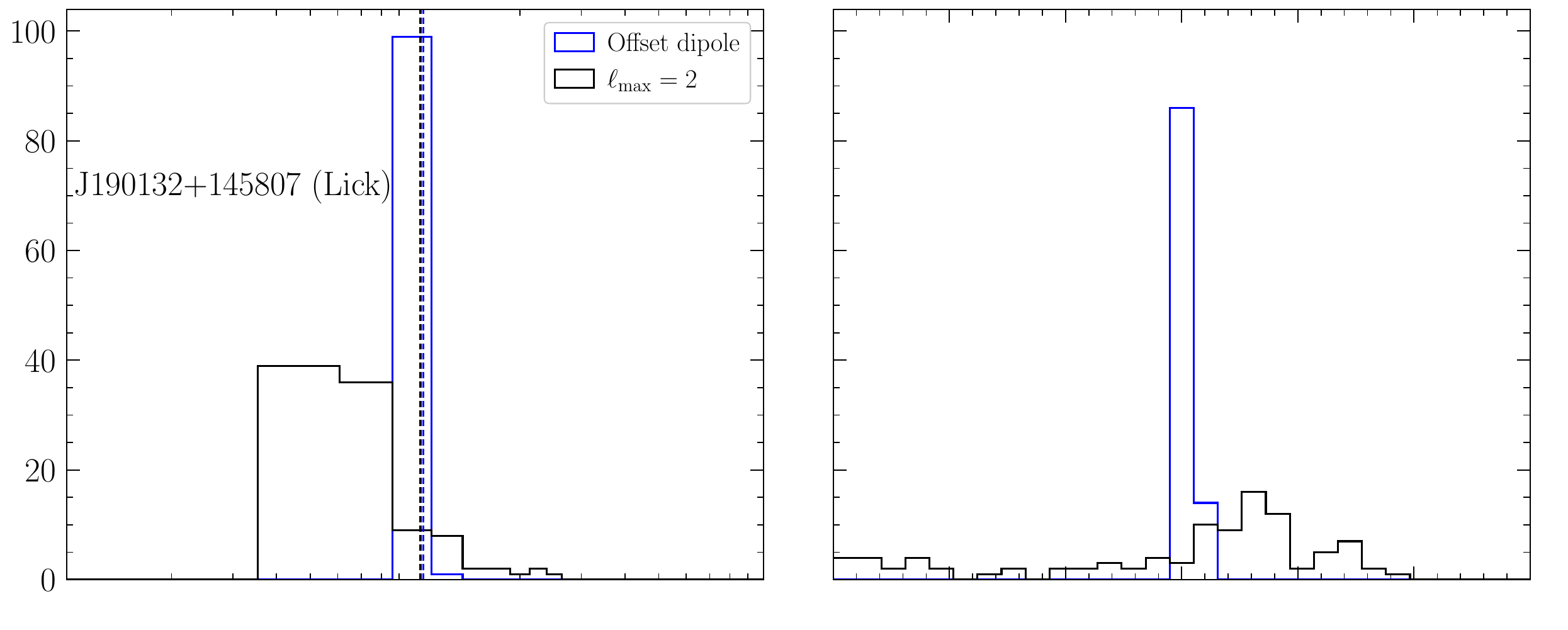} 
    \includegraphics[width=1.0\textwidth]{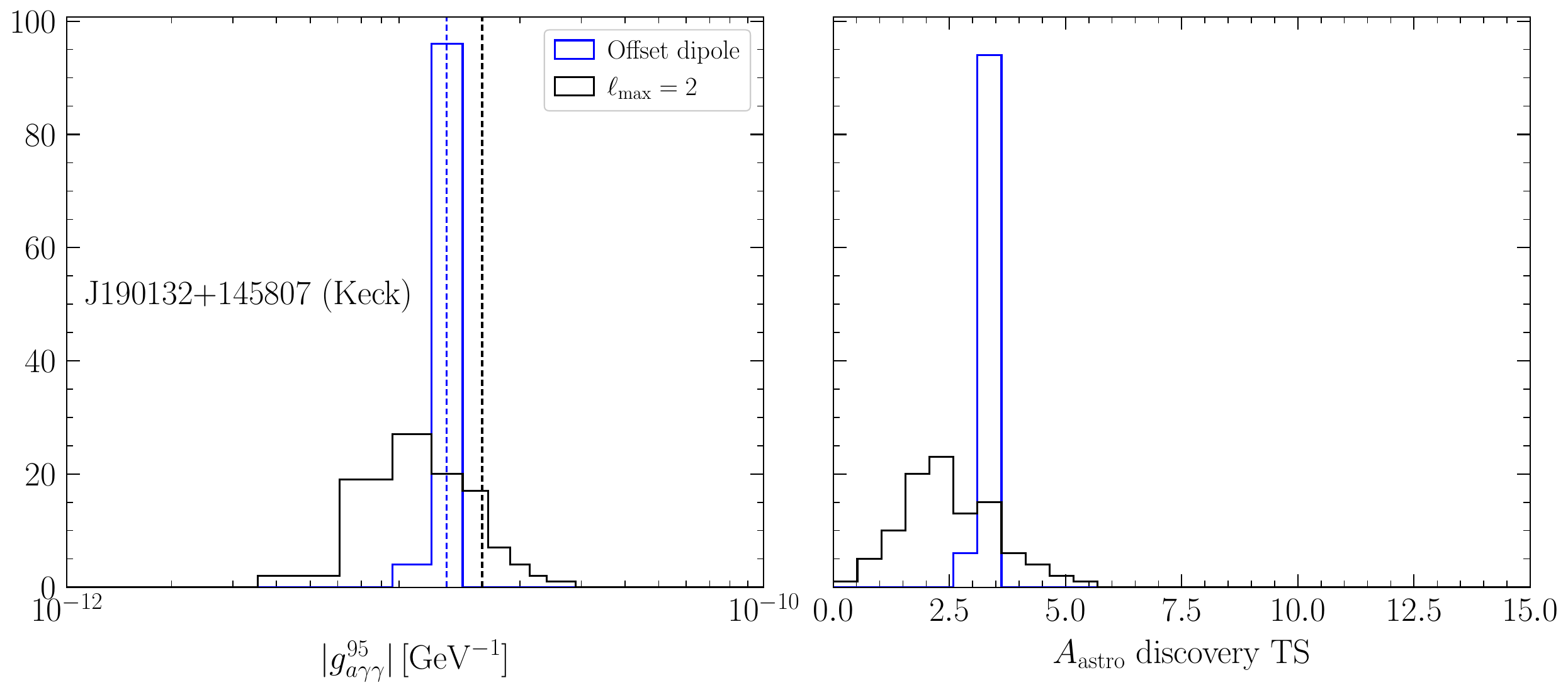} 
    \caption{(\textit{Upper panels}) As in Fig.~\ref{fig:Lick_j033320+000720_3panel_gayy_ul_95_dist_AVG}, but for the J190132+145807 Kast observation. (\textit{Lower panels}) The same, but for the J190132+145807 Keck/LRISp-ADC observation. }\label{fig:Keck_j190132+145807_3panel_gayy_ul_95_dist_AVG}
\end{figure}

In Fig.~\ref{fig:flux_fit_j190132+145807_v2}, we show the hydrogen transition lines used to assess the surface magnetic field, the flux data, and the linear polarization data taken from our LRISp-ADC (left) and Kast (right) observations. The magnetic field surface distributions we find are consistent not only between the offset dipole and harmonic field models but also between the two observations. Our findings are consistent with those of~\cite{Caiazzo:2021xkk}, which suggested field strengths $\sim600$--900 MG over the surface (note that this reference did not  fit magnetic atmosphere models to the data). In Figs.~\ref{fig:posterior_Keck_j190132+145807_ell_2},~\ref{fig:posterior_Lick_j190132+145807_ell_2},~\ref{fig:posterior_Lick_J190132_offset_dipole}, and~\ref{fig:posterior_Keck_J190132_offset_dipole}, we show the posterior distributions of the harmonic and offset dipole magnetic field model parameters fit to the LRISp-ADC and Kast flux data, respectively. 
For each model, in Fig.~\ref{fig:Keck_j190132+145807_3panel_gayy_ul_95_dist_AVG}, we show on the left the distribution of 95\% upper limits $|g_{a\gamma\gamma}^{95}|$ on the axion-photon coupling, and the discovery TS associated with the astrophysical model, over a sample of $100$ draws from the posterior distributions. The two models give broadly consistent limits, indicating the level of robustness of the result with respect to the magnetic field modeling.  We observe mild $\sim2\sigma$ evidence for astrophysical polarization associated with the MWD in the LRISp-ADC data. We see more significant evidence for astrophysical polarization in Kast, although we caution against overinterpreting this result. This observation was taken in poor observing conditions, on a cloudy night through a large airmass.

\section{Extended Details for the SDSS J002129+150223 Analysis}
\label{SM:other_2}

\begin{figure}[ht]
    \centering
    \includegraphics[width=0.5\textwidth]{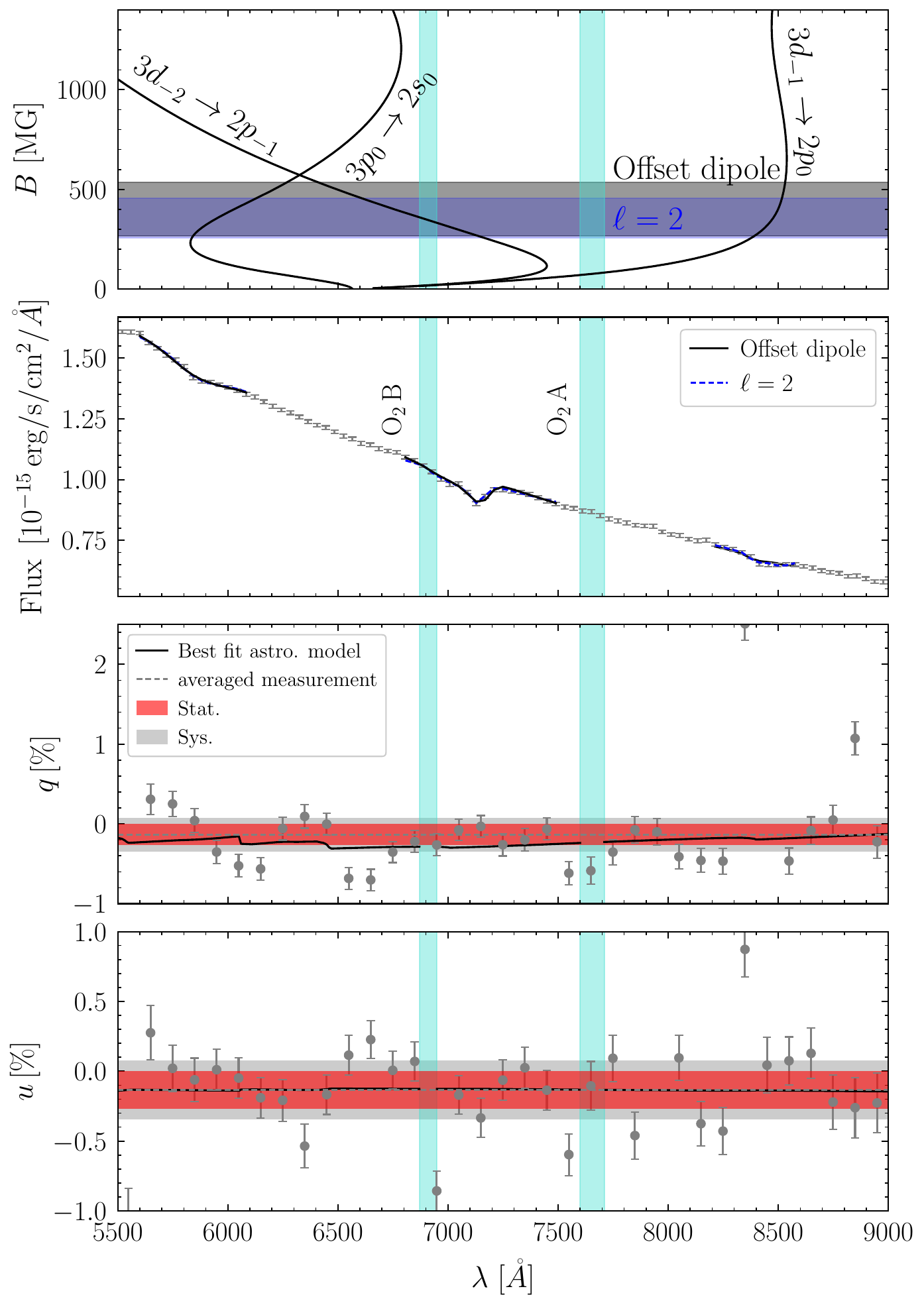}
    \caption{As in Fig.~\ref{fig:flux_fit_J033320+00072} but for the Keck/LRISp-ADC observation of J002129+150223.
    }. 
    \label{fig:flux_fit_j002129+150223_v2}
\end{figure}

\begin{figure*}[ht]
    \centering
    \includegraphics[width=1.0\textwidth]{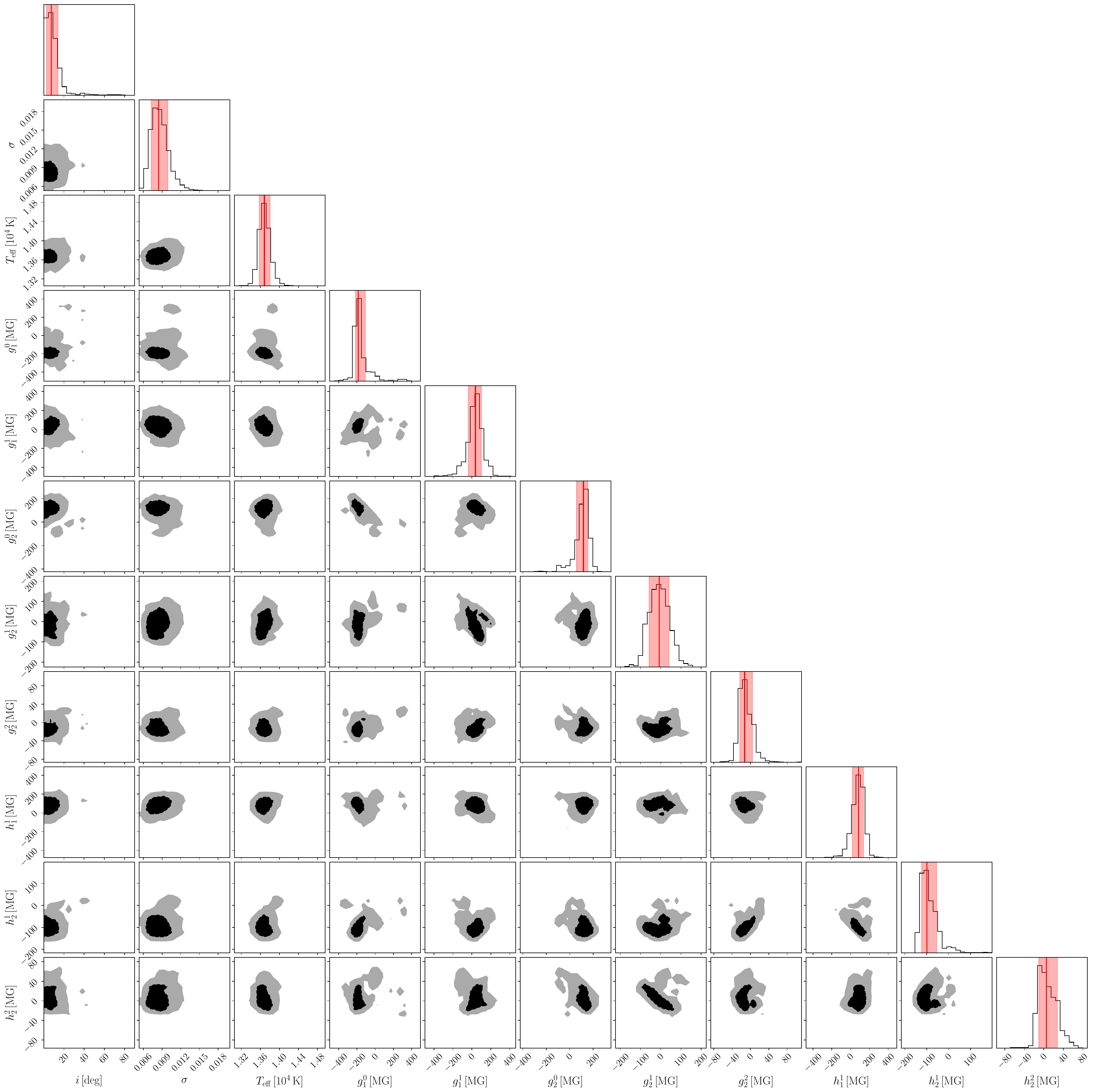} 
    \caption{As in Fig.~\ref{fig:posterior_Lick_J033320+00072_ell_2} but for the Keck/LRISp-ADC measurement of J002129+150223.  
    }
    \label{fig:posterior_Keck_j002129+150223_ell_2}
\end{figure*}

\begin{figure*}[ht]
    \centering
    \includegraphics[width=0.8\textwidth]{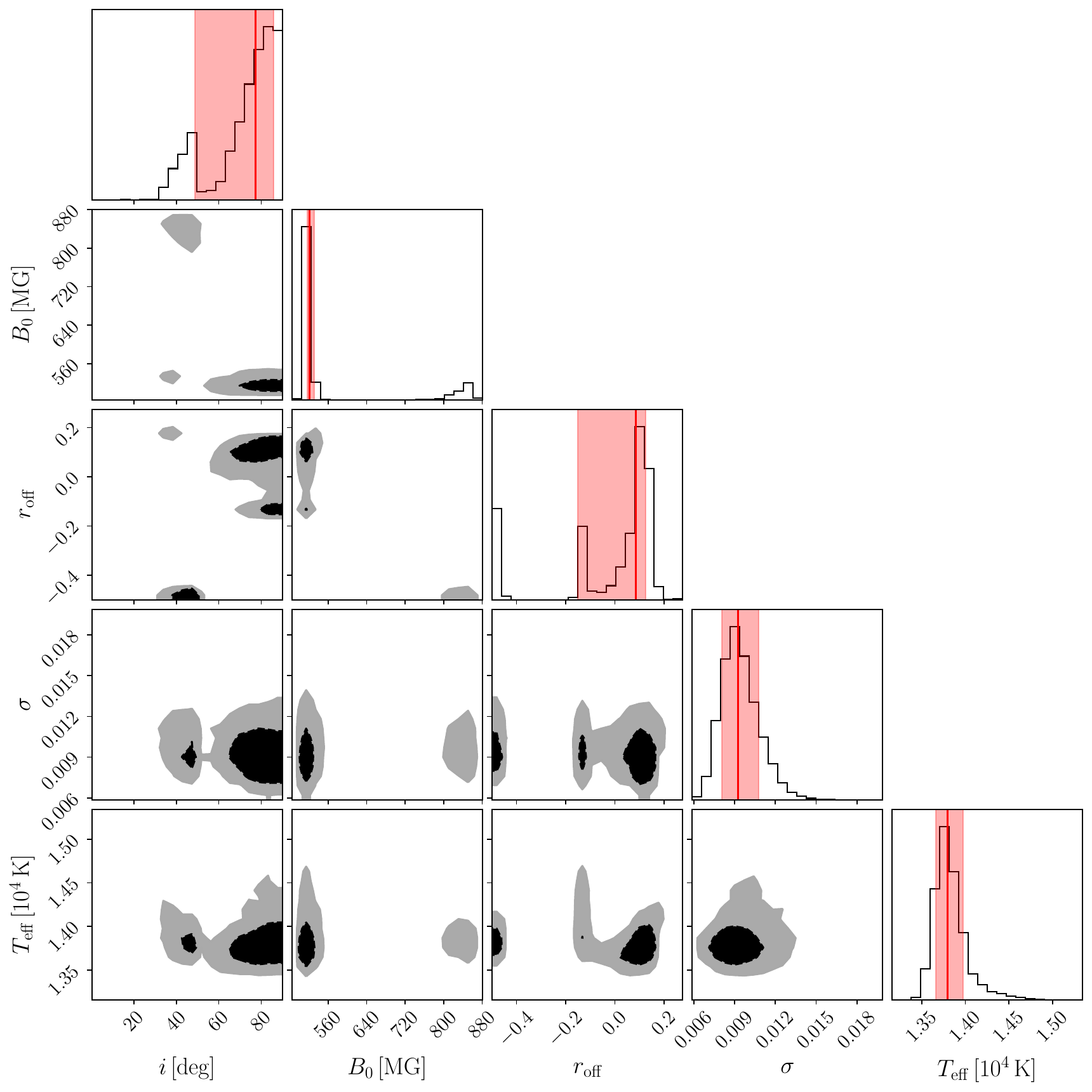} 
    \caption{The posterior distribution in the offset-dipole model for the Keck/LRISp-ADC observation of J002129+150223. 
    }
\label{fig:posterior_Keck_J002129+150223_offset_dipole}
\end{figure*}

\begin{figure}[h]
    \centering
    \includegraphics[width=1.0\textwidth]{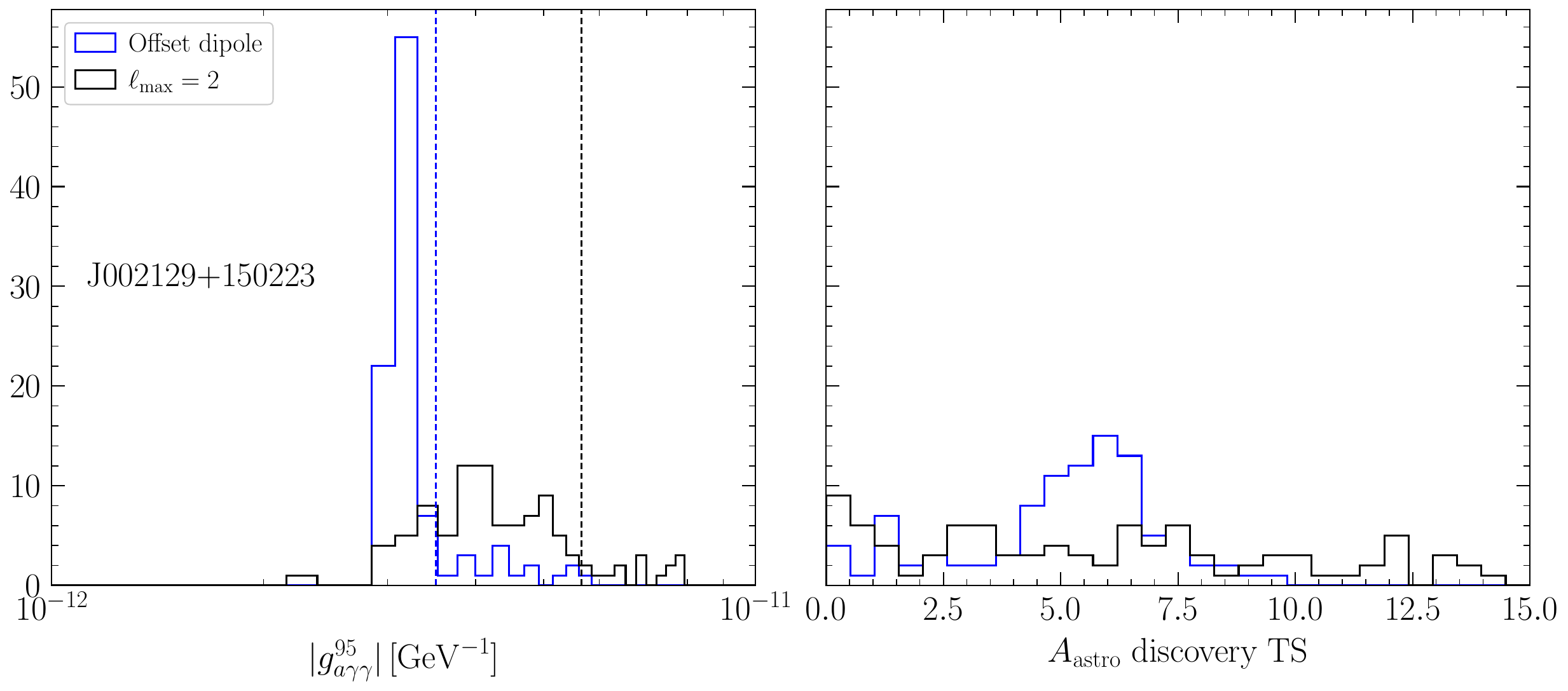} 
    \caption{As in Fig.~\ref{fig:Lick_j033320+000720_3panel_gayy_ul_95_dist_AVG}, but for the J002129 Keck/LRISp-ADC observation.}
    \label{fig:Keck_j002129+150223_3panel_gayy_ul_95_dist_AVG}
\end{figure}

Our analysis results for SDSS J002129+150223, which we observe with Keck/LRISp-ADC, are summarized in Figs.~\ref{fig:flux_fit_j002129+150223_v2}, \ref{fig:posterior_Keck_j002129+150223_ell_2}, \ref{fig:posterior_Keck_J002129+150223_offset_dipole}, and \ref{fig:Keck_j002129+150223_3panel_gayy_ul_95_dist_AVG}.  
The magnetic field geometry we constrain is consistent with Ref.~\cite{Kulebi:2009kp}, which found for a centered dipole model  $B_0 = 530 \pm 60$ MG and $\iota= 53^\circ \pm 26^\circ$, and for an offset-dipole $B_0 = 530 \pm 100\, \mathrm{MG}$,  $r_\mathrm{off}= 0.16 \pm 0.08$, and  $\iota =28^\circ \pm 52^\circ$. Note that the harmonic model gives a weakest upper limit $|g^{95}_{a\gamma\gamma}|$ at $1\sigma$  of $ 5.7\times 10^{-12} \, \mathrm{GeV}^{-1}$, whereas taking the smallest allowed values at $1\sigma$ of the centered-dipole fit parameters of Ref.~\cite{Kulebi:2009kp} gives  $|g^{95}_{a\gamma\gamma}|=7.45 \times 10^{-12}\, \mathrm{GeV}^{-1}$ for $m_a=0$. We find marginal evidence for astrophysical polarization under some of the magnetic field distributions.

\section{Extended Details for the SDSS J100356+053825 Analysis}
\label{SM:other_3}

\begin{figure}[ht]
    \centering
    \includegraphics[width=0.5\textwidth]{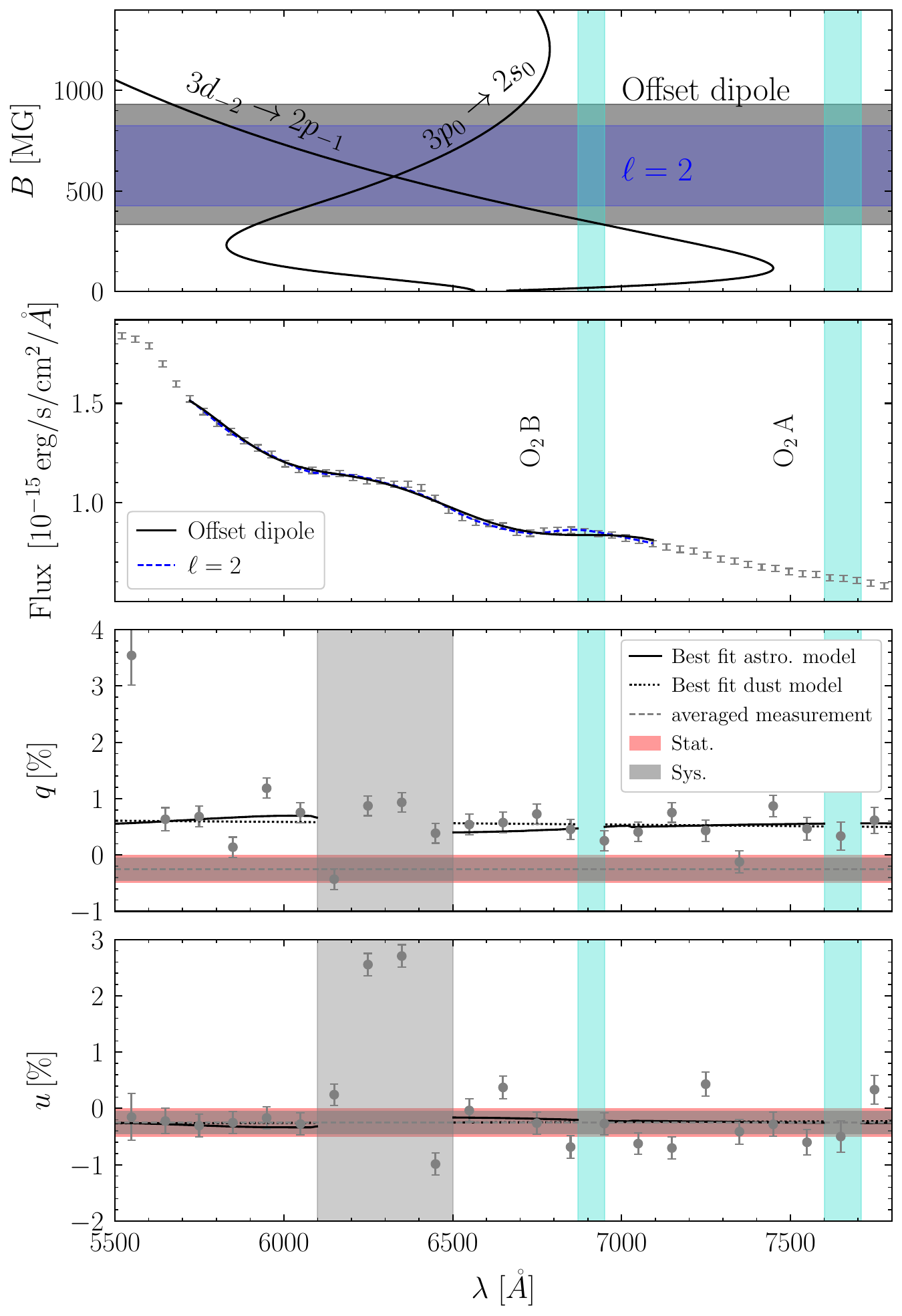}
    \caption{As in Fig.~\ref{fig:flux_fit_J033320+00072} but for the Keck/LRISp-ADC observation of J100356+053825. In addition to the oxygen absorption bands, in the analysis of the Stokes parameters we also mask wavelengths between 6100 and 6500 $\mathring{A}$, which likely feature spurious polarization due to a cosmic ray hitting the detector. We also show the best fit polarization model for interstellar dust (plus systematic offset).
    }
    \label{fig:flux_fit_j100356+053825_v2}
\end{figure}

\begin{figure*}[ht]
    \centering
    \includegraphics[width=1.0\textwidth]{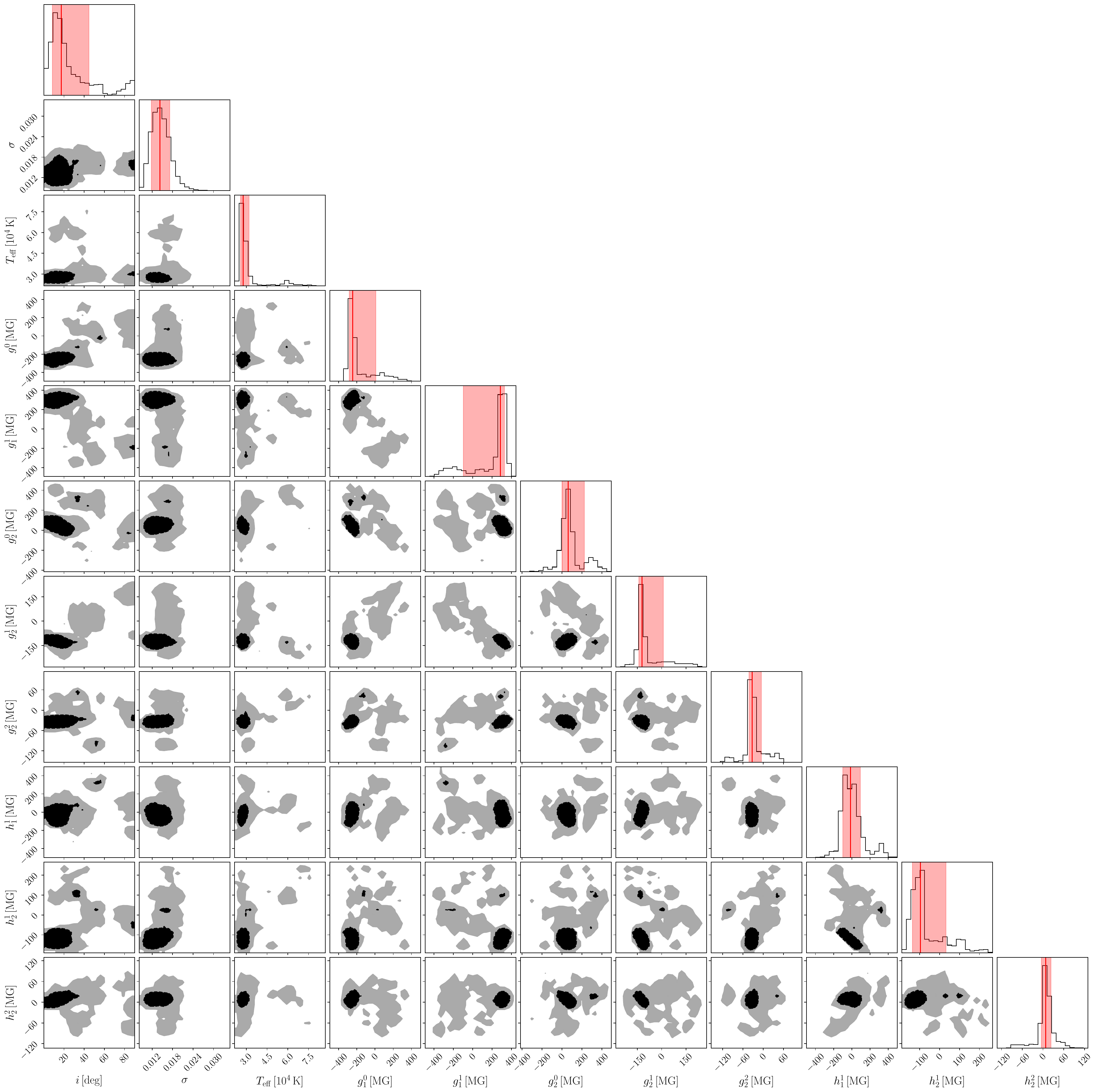} 
    \caption{As in Fig.~\ref{fig:posterior_Lick_J033320+00072_ell_2} but for the Keck/LRISp-ADC measurement of J100356.32+053825.6.  
    }
    \label{fig:posterior_Keck_J100356.32+053825.6_ell_2}
\end{figure*}

\begin{figure*}[ht]
    \centering
    \includegraphics[width=0.8\textwidth]{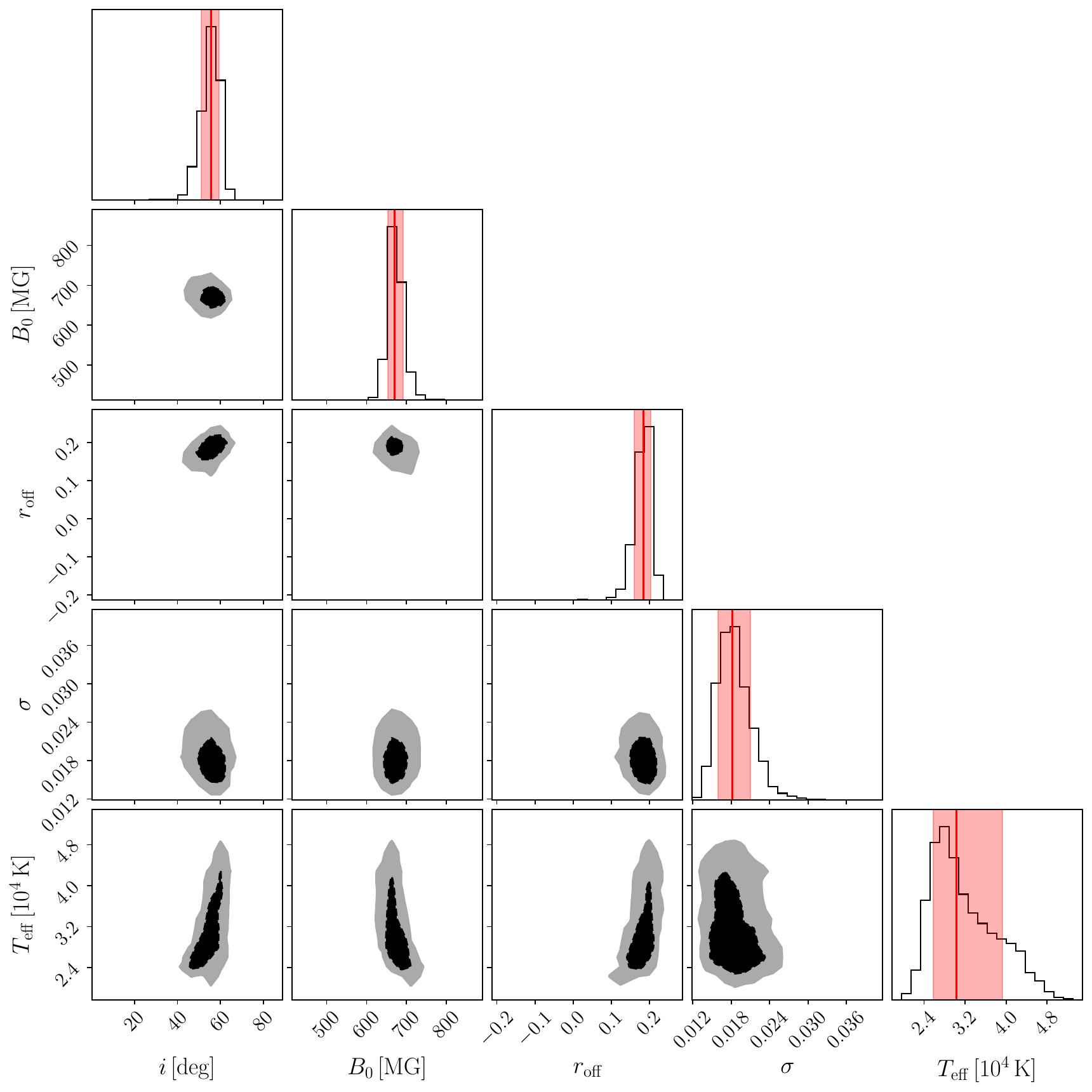} 
    \caption{The posterior distribution in the offset-dipole model for the Keck/LRISp-ADC observation of J100356.32+053825.6. 
    }
\label{fig:posterior_Keck_J100356.32+053825.6_offset_dipole}
\end{figure*}

\begin{figure}[h]
    \centering
    \includegraphics[width=1.0\textwidth]{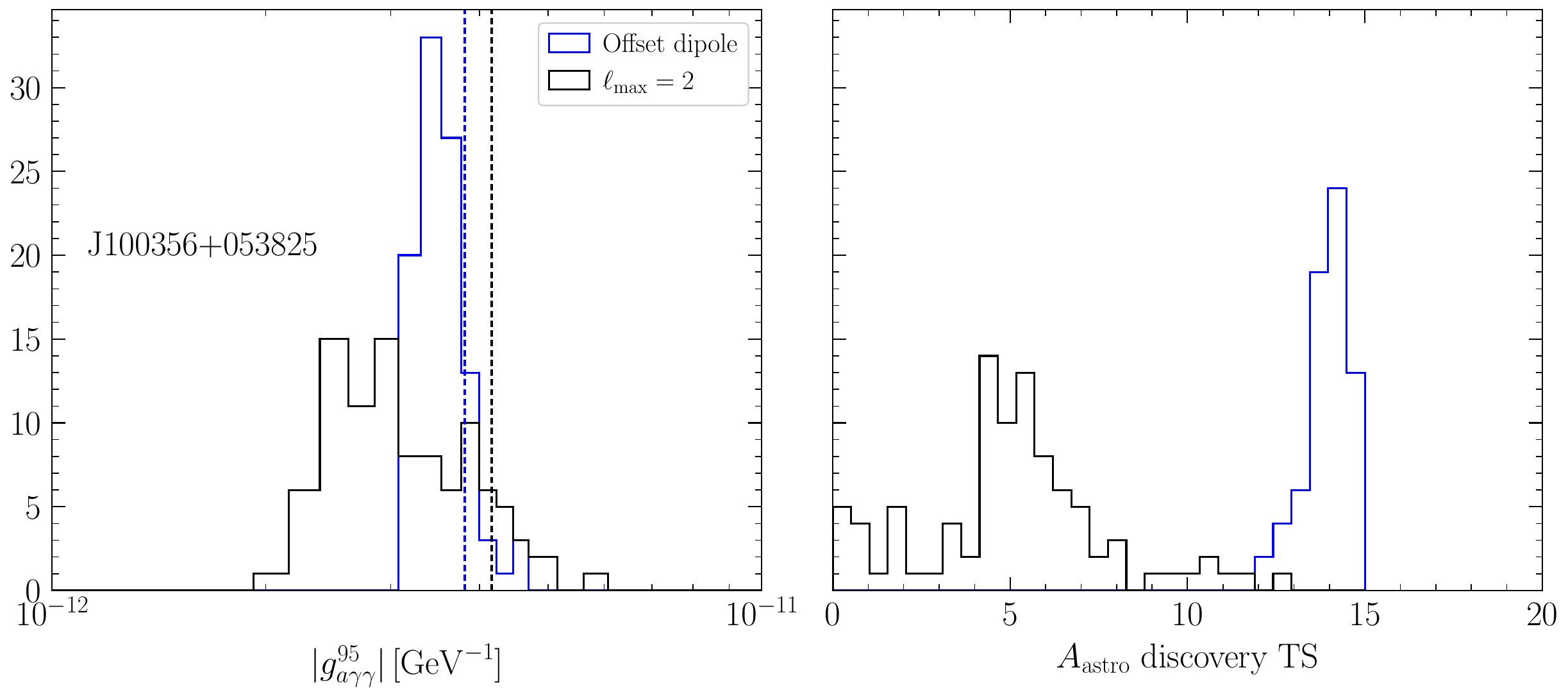} 
    \caption{As in Fig.~\ref{fig:Lick_j033320+000720_3panel_gayy_ul_95_dist_AVG}, but for the J100356.32+053825.6 Keck/LRISp-ADC observation.}\label{fig:Keck_j100356+053825_3panel_gayy_ul_95_dist_AVG}
\end{figure}

Our analysis of J100356 is the same as for the other MWDs observed with the Keck/LRISp-ADC, except that this observation did not observe wavelengths larger than 7800\,\AA.
The results are shown in Figs.~\ref{fig:flux_fit_j100356+053825_v2}, 
\ref{fig:posterior_Keck_J100356.32+053825.6_ell_2}, \ref{fig:posterior_Keck_J100356.32+053825.6_offset_dipole}, and \ref{fig:Keck_j100356+053825_3panel_gayy_ul_95_dist_AVG}.
Our characterization of the magnetic field geometry is consistent with Ref.~\cite{Kulebi:2009kp}, which found in the offset-dipole model
$B_0 = 668 \pm 131\, \mathrm{MG}$,  $r_\mathrm{off}=  0.08 \pm 0.05$,  $\iota =37^\circ \pm 23^\circ$, 
and in the centered dipole model  $B_0 = 670 \pm 120$\,MG, $\iota= 41^\circ \pm 35^\circ$. Our fiducial upper limit $|g^{95}_{a\gamma\gamma}|$ for the harmonic model is $ 4.2 \times 10^{-12} \, \mathrm{GeV}^{-1}$, which may be compared to the conservative upper limit resulting from the centered-dipole fit of Ref.~\cite{Kulebi:2009kp}, which we compute to be  $|g^{95}_{a\gamma\gamma}|=3.4 \times 10^{-11}\, \mathrm{GeV}^{-1}$ for $m_a=0$.

This MWD shows evidence for nonzero polarization, although it is not obviously arising from photon propagation through the MWD atmosphere. In particular, the bump in $u$ at $\sim 6300 \mathring{A}$ is likely due to a cosmic ray hitting the trace during data collection. For this reason when analyzing the Stokes parameter data, we mask wavelengths between 6100 $\mathring{A}$ and 6500 $\mathring{A}$. Additionally, of our entire sample this MWD has the largest extinction, and therefore is the most susceptible to continuum polarization arising from interstellar medium dust grains. Indeed, we expect $p_{\rm dust} \lesssim 0.5\% \pm 0.2\%$, in qualitative agreement with the observation. Dust leads to linear polarization spectra with wavelength-dependence~\cite{Andersson2015}
\begin{equation}
    L_p(\lambda)=p_{\rm max} \exp(-K \ln^2(\lambda_{\rm max}/\lambda))
\end{equation}
where $p_{\rm max}$ is the maximum polarization, set by the MWD extinction, observed at $\lambda_{\rm max}$, set by the typical size of the dust grains. Typical dust grains give $\lambda_{\rm max} \in [4000,7000]$\AA. $K$ is a constant which is observed to be linearly correlated with $\lambda_{\rm max}$, $K \approx 1.66 (\lambda_{\rm max}/\mu \mathrm{m})$~\cite{1992ApJ...386..562W}. We fit this model to our data, and we show our best fit in Fig.~\ref{fig:flux_fit_j100356+053825_v2}. Our data prefer $p_{\rm max} \sim 0.6\%$, $\lambda_{\rm max} \sim 4100$\,\AA, in good agreement with the extinction measurements at this MWD (see Tab.~\ref{tab:obslog}). We therefore expect that the observed polarization may be due to dust rather than intrinsic to the MWD.

\end{document}